\documentclass[iop]{emulateapj}
\usepackage{graphicx,epsfig,amssymb,color,lineno}

\usepackage{natbib}
\usepackage{amsmath}

\slugcomment{\footnotesize Accepted for publication in ApJ}
\shorttitle{UV Extinction and Luminosity Function at $z\sim2$}
\shortauthors{Alavi et al.}

\begin{document}

\title{Ultra-faint Ultraviolet Galaxies at $z\sim2$ Behind the Lensing Cluster Abell 1689: the Luminosity Function, Dust Extinction and Star Formation Rate Density\footnotemark[1]} \footnotetext[1]{Some of the data presented herein were obtained at the W.M. Keck Observatory, which is operated as a scientific partnership among the California Institute of Technology, the University of California and the National Aeronautics and Space Administration. The Observatory was made possible by the generous financial support of the W.M. Keck Foundation.}

\author{\sc Anahita Alavi, Brian Siana\altaffilmark{2}, Johan Richard\altaffilmark{3}, Daniel P. Stark\altaffilmark{4}, Claudia Scarlata\altaffilmark{5}, Harry I. Teplitz\altaffilmark{6}, William R. Freeman\altaffilmark{2}, Alberto Dominguez\altaffilmark{2}, Marc Rafelski \altaffilmark{6}, Brant Robertson\altaffilmark{4}, Lisa Kewley\altaffilmark{7}}

\altaffiltext{2}{Department of Physics and Astronomy, University of California, Riverside, CA 92521, USA} 
\altaffiltext{3}{Centre de Recherche Astrophysique de Lyon, Universit\'e Lyon 1, 9 Avenue Charles Andr\'e, F-69561 Saint Genis Laval Cedex, France}
\altaffiltext{4}{Department of Astronomy, Steward Observatory, University of Arizona, 933 North Cherry Avenue, Rm N204, Tucson, AZ, 85721}
\altaffiltext{5}{Minnesota Institute for Astrophysics, University of Minnesota, Minneapolis, MN 55455, USA}
\altaffiltext{6}{Infrared Processing and Analysis Center, Caltech, Pasadena, CA 91125, USA}
\altaffiltext{7}{Research School of Astronomy and Astrophysics, The Australian National University, Cotter Road, Weston Creek, ACT 2611}

\begin{abstract}
We have obtained deep ultraviolet imaging of the lensing cluster Abell 1689 with the WFC3/UVIS camera onboard the {\it Hubble Space Telescope} ({\it HST}) in the F275W (30 orbits) and F336W (4 orbits) filters.  These images are used to identify $z\sim2$ star-forming galaxies via their Lyman break, in the same manner that galaxies are typically selected at $z \geq 3$.  Because of the unprecedented depth of the images and the large magnification provided by the lensing cluster, we detect galaxies 100$\times$ fainter than previous surveys at this redshift. After removing all multiple images, we have 58 galaxies in our sample between $-19.5<M_{1500}<-13$ AB mag. Because the mass distribution of Abell 1689 is well constrained, we are able to calculate the intrinsic sensitivity of the observations as a function of source plane position, allowing for accurate determinations of effective volume as a function of luminosity.  We fit the faint-end slope of the luminosity function to be $\alpha=-1.74\pm0.08$, which is consistent with the values obtained for $2.5<z<6$.  Notably, there is no turnover in the luminosity function down to $M_{1500}=-13$ AB mag.  We fit the UV spectral slopes with photometry from existing {\it Hubble} optical imaging. The observed trend of increasingly redder slopes with luminosity at higher redshifts is observed in our sample, but with redder slopes at all luminosities and average reddening of $<E(B-V)>=0.15$ mag. We assume the stars in these galaxies are metal poor ($0.2\ Z_{\sun}$) compared to their brighter counterparts ($Z_{\sun}$), resulting in bluer assumed intrinsic UV slopes and larger derived values for dust extinction. The total UV luminosity density at $z\sim2$ is $4.31\substack{+0.68\\-0.60}\times10^{26}$ erg s$^{-1}$ Hz$^{-1}$ Mpc$^{-3}$, more than 70\% of which is emitted by galaxies in the luminosity range of our sample. Finally, we determine the global star formation rate density from UV-selected galaxies at $z\sim2$ (assuming a constant dust extinction correction of 4.2 over all luminosities and a Kroupa IMF) of $0.148\substack{+0.023\\-0.020}$ M$_{\sun}$ yr$^{-1}$ Mpc$^{-3}$, significantly higher than previous determinations because of the additional population of fainter galaxies and the larger dust correction factors.

\end{abstract}


\section{Introduction}

A primary goal of observational cosmology has been to obtain a complete census of star formation at all epochs \citep[e.g.,][]{mad96,lil96}. One effective method of selecting star-forming galaxies is to identify a ``Lyman break" in the ultraviolet (UV) continuum caused by Lyman line and continuum absorption from hydrogen in the stellar atmospheres, interstellar medium, and the intergalactic medium \citep{ste99}. This technique was first used to select the star-forming galaxies as U-band ``dropouts" at z$\sim$3 \citep{ste99}, and has since been widely adopted to select Lyman break galaxies (LBGs) at higher redshifts \citep[e.g.,][]{bun04,saw06,yan06,bou07,bou09,red09,oes10b,bun10,yan10,hat12}.  

Although the peak epoch of star formation is likely at lower redshifts \citep[$1.5<z<3$, e.g.,][]{ly09,red09}, it has been impossible to select galaxies at these redshifts via their Lyman break without imaging at wavelengths below the atmospheric limit, $\lambda < 3100$ \AA.  Other selection methods such as the BM/BX \citep{ade04} and BzK \citep{dad04} criteria have been used, but unambiguous comparison with Lyman break selected galaxies at $z\ge3$ is difficult because the selection effects are different.    

The new Wide--Field Camera 3 (WFC3) on the {\it Hubble Space Telescope} has an ultraviolet/optical channel (UVIS) that increases survey efficiency (area $\times$ throughput) by more than an order of magnitude compared with WFPC2.  With the first ultraviolet images in the Early Release Survey (ERS) \citep{win11}, star-forming galaxies were selected via the Lyman break technique at $1<z<3$ \citep{oes10a,hat10}.  These studies were limited by the depth of the shallow ultraviolet imaging and could only detect galaxies with absolute UV magnitude (typically measured at 1500 \AA, $M_{1500}$) brighter than $-19$ AB-mag, making it impossible to accurately constrain the faint-end slope of the UV luminosity function (LF).

Rest-frame ultraviolet light is a strong tracer of unobscured star formation, so the UV luminosity function can be used to determine the relative contribution of faint and bright galaxies to the total star formation rate density (SFRD). The UV LF has been studied widely at this active star forming era \citep{arn05,red09, hat10, oes10a}. 

There are two difficulties in determining a complete census of star formation. First, the ultraviolet light emitted by bolometrically luminous galaxies is significantly attenuated by dust, which causes UV-selected samples to be incomplete at the bright end.  However, these galaxies can be identified by their far-infrared emission \citep{red06,mag11}. Second, there may be a large population of faint star-forming galaxies beyond the detection limit of the deepest surveys (typically $M_{1500}<-18$ AB-mag).  

This population of faint ultraviolet galaxies may contribute significantly to the global star formation density and the ionizing background \citep{nes12} at $z>2$.  Recent searches at z=3 \citep{nes11} have revealed that faint star-forming galaxies on average have larger Lyman continuum escape fractions relative to the brighter galaxies. This result proves the importance of these feeble objects in maintaining the ionizing emissivity especially at the peak of star formation activity.
 In addition to the crucial role of the faint galaxies at lower redshifts, it is believed that low luminosity galaxies at $z>7$ are likely the primary sources of ionizing photons that caused the reionization of the intergalactic medium \citep{kuh12,rob13}. Studying an analogous population at lower redshifts ($z\sim2$) provides a clearer picture of ultra-faint populations at high redshifts.

One way to efficiently detect and study these faint galaxies is to use foreground massive systems as lenses to magnify background galaxies.  This strong gravitational lensing conserves surface brightness while spreading out a galaxies' light over a larger area and magnifying it.  Over the last decade, this has been used to study individual lensed galaxies in great detail \citep[e.g.,][]{pet02,sia08b,sia09,sta08, jon10,yua13}. 

When using strong lensing to identify large numbers of faint galaxies, galaxy clusters are particularly useful as they highly magnify background galaxies over a large area \citep{nar84}.  The regions amplified by a higher factor have smaller source-plane area, so the benefit of magnification is offset by reduced sample volume. Therefore the total number of candidate galaxies behind a cluster could be either smaller or larger than a field survey of the same depth and area, due to the competition between these two variables. Determining the ratio of the total number of galaxies found by using cluster lensing relative to the field studies, depends on the effective slope of the luminosity function \citep{bro95,bou09}. If the effective slope of the luminosity function, -$d$ log $\phi$ /$d$ log $L$, is greater than unity then a survey behind a lensing cluster will find more objects than the same survey in the field. Lensing clusters have been used to identify very high redshift objects \citep[e.g.,][]{kne04,ega05,sta07,bra08,ric08,bou09,ric11,bou12a,zhe12,zit12,coe13} because the luminosities probed are on the steep, bright end of the luminosity function, resulting in larger samples than field surveys. Historically, surveys have not used lensing clusters to identify galaxies significantly fainter than $L^*$ because it finds fewer galaxies. However, our primary concern is not the {\it number} of galaxies found, but the intrinsic {\it luminosity} of the galaxies. The average luminosity of galaxies found behind a lensing cluster will be significantly lower than surveys in the field. Because we are interested in finding galaxies that are undetected in our deepest field surveys, we chose to survey faint star-forming galaxies behind massive clusters. Once these ultra-faint galaxies are identified, the lensing will allow detailed study as they are highly magnified and the light is spread over many more resolution elements. This strategy of surveying large numbers of background galaxies with deep observations of lensing clusters will soon be adopted with deep {\it Hubble} imaging of the Frontier Fields beginning in Cycle 21.  

In this paper we use the WFC3/UVIS channel to look for faint star-forming galaxies located behind the massive cluster Abell 1689. This cluster has the most constrained cluster mass model due to the large number of confirmed multiply imaged systems (43), of which 24 have spectroscopic redshifts \citep{ lim07,  coe10}.  This mass model gives us a precise estimation of magnification factor over the total area. The high magnification area for this cluster is well-matched with the WFC3/UVIS field of view. Abell 1689 has been observed extensively with  {\it HST} in the optical with ACS/WFC (F475W, F625W, F775W, F814W, F850LP) and the near-IR with WFC3/IR (F105W, F125W, F140W, F160W), {\it Spitzer}  IRAC and MIPS as well as  {\it Herschel} PACS and SPIRE.

The outline of this paper is as follows. In Section  \ref{sec:observation}, we describe the observations and data reduction. The selection technique is given in Section \ref{sec:color selection}. In Section \ref{sec:completeness} we discuss the details of the completeness simulation as well as the Monte Carlo simulation used for IGM opacity. The redshift distribution of the sample is explained in this section. We explain the maximum likelihood method used for estimating the rest-frame UV luminosity function parameters in Section \ref{sec:LF}. In Section \ref{sec:dust} we determine the dust content of the selected Lyman break galaxies. In Section \ref{sec:discussion}, we compare our final results for the UV luminosity density and evolution of the faint-end slope and dust extinction with other results in the literature. We also briefly discuss about the effect of the intracluster dust. Finally, we present a short summary in Section \ref{sec:result}.

All distances and volumes are in comoving coordinates. All magnitudes are given in the AB system \citep{oke83}. We assume $\Omega_{M}=0.3$, $\Omega_{\Lambda}=0.7$ and $H_{ 0}=70$ km s$^{-1}$ Mpc$^{-1}$.

\section{Observations}
\label{sec:observation}
 We used the WFC3 UVIS channel to obtain images in F275W (30 orbits) and F336W (4 orbits) as part of {\it HST} Program ID 12201 (PI: B. Siana) between December, 2010 and March, 2011.  We used long exposure times, half an orbit in length (1310s each) in order to minimize the total read noise, as this is the dominant source of noise in near-UV imaging with {\it Hubble}. The data were obtained in two orbit visits during which we performed the standard UVIS dither pattern, WFC3-UVIS-DITHER-BOX.  Each dither pattern was offset up to $\pm3"$ from the central pointing to place different pixels on the same objects and to fully cover the UVIS chip gap. The 5$\sigma$ depths measured in a 0.2$\arcsec$ radius aperture are given in Table \ref{tab:depth}.  

\begin{deluxetable*}{ccccccc}
\tablecaption{WFC3/UVIS and ACS/WFC Magnitude Limits}
\tablewidth{0pt}
\tablehead{\colhead{Filter} & \colhead{F275W} & \colhead{F336W} & \colhead{F475W} & \colhead{F625W} &  \colhead{F775W} & \colhead{F850LP}}
\startdata
Magnitude Limit\tablenotemark{a}& 28.7 & 27.90 & 28.55& 28.29 & 28.17 & 27.80
\enddata
\tablenotetext{a}{5$\sigma$ limit in a 0.2$\arcsec$ radius aperture}
\label{tab:depth}
\end{deluxetable*}

In order to identify the LBGs at $z\sim2$, we also used the existing {\it HST}/ACS images in optical bands (F475W, F625W, F775W, F850LP; PID 9289, PI: H. Ford). Table \ref{tab:depth} shows the limiting magnitudes of these observations.  The overlapping area between the ACS/WFC and WFC3/UVIS fields of view, after subtracting the areas contaminated by the bright cluster galaxies, covers 6.56 arcmin$^{2}$ in the image plane.

\subsection{Data Reduction}
The calibrated, flat fielded WFC3/UVIS and ACS/WFC images were processed with {\it MultiDrizzle} \citep{koe03}, part of the STSDAS/DITHER IRAF package \citep{tod86}. The initial drizzled images were registered to SDSS images in order to compute the shift file required for astrometric correction. Because each visit has a slightly different pointing and small differences in rotation and alignment, we make a shift file for each visit to project all the images to the same astrometric reference grid.

The F336W image was aligned with the SDSS g$'$-band image with a precision of  0.1\arcsec, using unsaturated stars and compact objects. We chose to use the UVIS F336W image to align with SDSS because many of the stars were not saturated at these wavelengths and the galaxies were generally more compact, resulting in more precise alignment.

The other {\it HST} images were matched to F336W to achieve astrometric registration with the SDSS reference frame. The relative alignment of images was done with a precision of 0.1 pixels (0.004\arcsec) because of its importance in doing matched-aperture photometry in all filters. The shift files were created by running the {\it geomap} task in IRAF. These shift files were then used as input to re-run {\it MultiDrizzle}.

The input images to the {\it MultiDrizzle} software were drizzled onto separate undistorted output frames which were combined later into a median image. The median image was transformed back (blotted) to the original distorted images in order to make the cosmic ray masks. The final output is a registered, undistorted and cosmic ray-rejected image. We also set {\it MultiDrizzle} to produce an inverse variance weight image to be used for computing uncertainties in the photometry. In order to do matched-aperture photometry for all wavelengths, we set the same output pixel sizes for WFC3/UVIS and ACS/WFC images to 0.04\arcsec. We set the pixfrac, fraction by which the input pixels are shrunk before being drizzled, to 0.8 for ACS/WFC and 1 for WFC3/UVIS images since these pixfrac values were well matched to our output pixel scale.

 The sources were detected in the ACS {\it B} band image (F475W) using {\it SExtractor} \citep{ber96}. The photometry was done by running {\it SExtractor} in dual image mode using the weight map RMS-MAP generated by {\it MultiDrizzle}. We used isophotal apertures with detection threshold 1.27$\sigma$ and minimum area 16 pixels. We ran {\it SExtractor} with high and low values for the de-blending minimum contrast parameter without changing other parameters in the {\it SExtractor} configuration file. We were able to detect very faint galaxies as separate objects in the catalog with very low DEBLEND-MINCONT=0.005 and then add them to the other catalog produced with the larger DEBLEND-MINCONT =0.13 parameter. This method is similar to hot and cold detections used in \citet{rix04}. All of the isophotal magnitudes are given in AB magnitudes by using the WFC3 and ACS photometric zero points provided by STScI.

The output  {\it SExtractor} uncertainties don't include the correlations between pixel counts that result from combing the input images with {\it MultiDrizzle}. Following \citet{cas00} paper, the correction factor ($\sqrt{F_{A}}$) which is approximately the ratio of uncorrelated noise to correlated error from {\it SExtractor}, can be estimated as below:   
  \begin{equation}
   \sqrt{F_{A}}=\begin{cases}
    \frac{s}{p}(1-\frac{1}{3}\frac{s}{p}) &  \text {$s < p$}\\
    1-\frac{1}{3}\frac{p}{s}                       &  \text{$s > p$}
    \end{cases}
   \end{equation}
where $p$ is the pixfrac and $s$ is the ratio of output pixel size to original pixel size.
  
We also supplemented our WFC3/UVIS and ACS/WFC images with WFC3/IR data, in order to compute the photometric redshifts. The higher resolution near-UV and optical images were convolved with Gaussians to degrade the resolution to that of the WFC3/IR images.  The photometry was then measured using matched apertures in all images.

\section{Color Selection and Sample}
\label{sec:color selection}
\begin{figure}
\epsscale{2.0}
\includegraphics[trim=1cm 0cm 0.5cm 0.5cm,clip=true,width=\columnwidth]{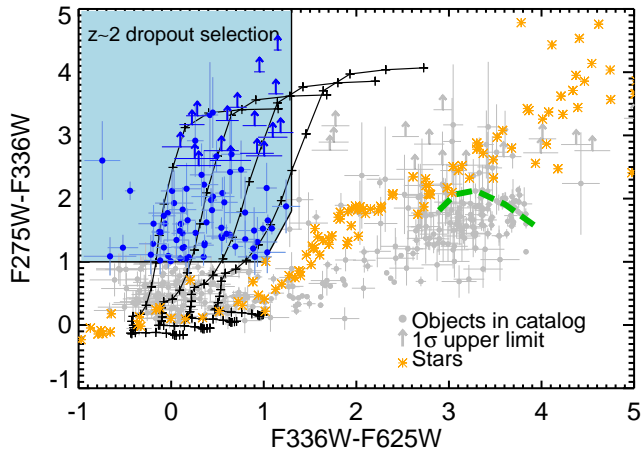}
\caption{Color selection of $z \sim2$ Lyman-break galaxies (F275W-dropouts). Grey dots are objects detected at greater than $5\sigma$ significance in F336W and F625W filters. Grey arrows represent $1\sigma$ upper limits. Black lines are tracks of star-forming galaxies that are dust obscured with E(B-V)=\{0, 0.1, 0.2, 0.3\} (in magnitudes). Orange asterisks are stars from \citet{pic98}. The blue shaded region is the region selected by the criteria given in Equation \ref{equ:color-selection}. The blue circles and blue arrows indicate our candidates. The green dashed line is the expected color track for lower redshift ($0.0<z<0.2$) elliptical galaxies \citep{col80}. Abell 1689 is at $z=0.18$.  The large number of redder galaxies (in both colors) are primarily galaxies in Abell 1689.}
{( A color version of this figure is available in the online journal)}
\label{fig:color}
\end{figure}

With our near-UV images we apply the Lyman break selection of $z\sim2$ galaxies, allowing for direct comparison with $z\ge3$ studies. The selection region is defined by the location of star-forming SEDs in color-color space. Our selection criteria, which are shown in Figure \ref{fig:color}, were found by running \citet{bru03} (hearafter BC03) models with constant star formation for 100 Myr, reddened by applying Calzetti attenuation curve \citep{cal00} with $E(B-V)=\{ 0, 0.1, 0.2, 0.3\}$ (in magnitudes)  and IGM obscuration from \citet{mad95}. The green dashed line is the track of lower redshift elliptical galaxies which is extended to $z=0.2$, as Abell 1689 is at $z=0.18$.

Our color selection criteria for selecting the F275W dropouts are:
\begin{displaymath}
F275W-F336W > 1
\end{displaymath}
\begin{displaymath}
F336W-F625W < 1.3
\end{displaymath}
\begin{displaymath}
F275W-F336W > 2.67 (F336W-F625W) - 1.67
\end{displaymath}
\begin{displaymath}
S/N(F336W) > 5 \hspace{0.5cm} , \hspace{0.5cm} S/N(F625W) > 5
\end{displaymath}
\begin{equation}
\label{equ:color-selection}
\end{equation}

We found 84 candidates using these criteria.  Four candidates were dismissed as they were fainter than the limiting magnitude in Table \ref{tab:depth} (F336W=27.9 mag and F625W=28.29 mag). Because of the strong lensing phenomenon, many of the candidates in our sample are multiply imaged. In order to compute the luminosity function, we must remove all but one of the multiple images from our catalog. We removed all of the previously known multiple images (12) except the brightest image in each system. In addition to these confirmed multiple images, the Abell 1689 mass model predicts all of the possible counterimages for each object as a function of redshift. Using these predictions, we found eight new multiple images by performing a visual inspection considering the photometric redshift of each object (A. Dominguez et al., in preparation, also see Subsection \ref{sebsec:incomp_cor}). We removed two objects with photometric redshifts less than 1.3. Our final sample consists of 58 $z\sim 2$ LBG candidates (see Section \ref{sec:completeness} ). 

The purity of our sample is quite high because of two main reasons. First, the possible contamination from other sources (e.g, stars) can be recognized easily, because these images are high resolution. Second, since these galaxies are UV dropouts, it is very unlikely to have any other break except Lyman break in the UV band. However the contamination from similar galaxies (LBGs) at slightly either higher or lower redshifts, is not negligible. We account for these possible contaminants in our completeness simulation (see Subsection \ref{sebsec:incomp_cor}).

\begin{figure}
\epsscale{2.0}
\includegraphics[trim=0cm 0cm 0cm 0cm,clip=true,width=\columnwidth]{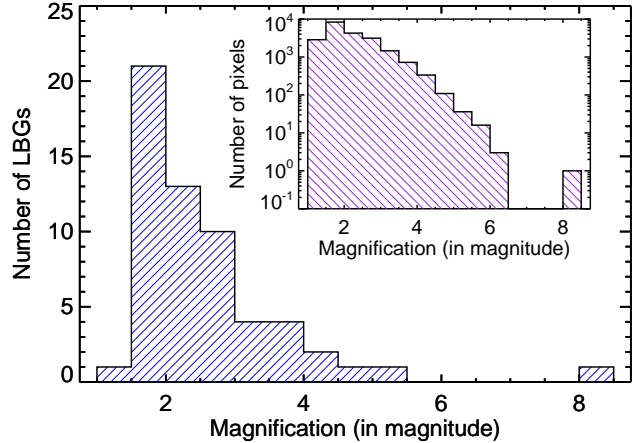}
\caption{The magnification distribution function of Lyman-break galaxies. The magnification values are from the cluster lens model assuming the spectroscopic redshifts for 12 objects (see Subsection \ref{sebsec:incomp_cor}) and an average redshift of $z=2$ (see Subsection \ref{subsec:redshift}) for the rest of the LBGs. The inset shows the distribution function of magnification values over all field of view pixels which are projected on the source plane at $z=2$.}
{( A color version of this figure is available in the online journal)}
\label{fig:magnification}
\end{figure}

The selected Lyman break galaxies have observed $B$-band magnitude down to $m_{F475W}< 27.5 $ mag, but are intrinsically fainter as they are all highly magnified. The lensing cluster mass model estimates the magnification value at each point of the image as a function of redshift. The magnification of each LBG was measured by assuming the spectroscopic redshifts for 12 out of 58 LBGs (see Subsection \ref{sebsec:incomp_cor}) and an average of $z=2$ (see Subsection \ref{sec:completeness}) for the rest of the sample. Figure \ref{fig:magnification} shows the distribution of magnification in magnitude units for all the UV-dropout galaxies. The fluxes are on average magnified by a factor of 10 (2.5 magnitude) and for some galaxies the magnification becomes very large (up to 8 magnitude). The distribution of magnification over the whole field of view in the source plane at $z=2$ is given in Figure \ref{fig:magnification} (inset), which shows all the pixels in this area are magnified by at least one magnitude.

\begin{figure}
\epsscale{2.0}
\includegraphics[trim=1cm 0cm 0.5cm 0.5cm,clip=true,width=\columnwidth]{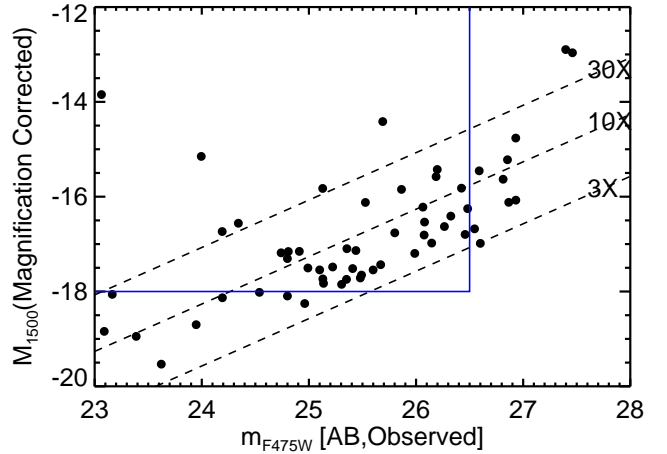}
\caption{The intrinsic absolute magnitude, corrected for magnification, is plotted vs. observed B-band magnitude. There are 39 LBGs in the blue box which denotes galaxies that are both intrinsically faint ($M_{1500}>-18$) {\it and} bright enough ($B < 26.5$) for follow-up ground-based spectroscopy. Dashed lines show constant magnifications (3, 10 and 30 in flux density units)  }
{( A color version of this figure is available in the online journal).}
\label{fig:M-m}
\end{figure}

Due to the large magnifications, the intrinsic absolute magnitudes measured at rest frame 1500 \AA, M$_{1500}$, probed in this study  go down to very faint magnitudes (M$_{1500} < -13$ mag), about 100 times fainter than previous studies (M$_{1500} < -18$ mag) at the same redshift \citep{red09,hat10,oes10a,saw06}. The intrinsic absolute magnitudes corrected for magnification (M$_{1500}$) versus observed B-band magnitudes (F475W) are plotted in Figure \ref{fig:M-m}. The Dashed lines represent the fixed magnifications of 3, 10 and 30 in flux density units. There are 39 galaxies within the blue box which are both intrinsically very faint (M$_{1500}$ $>-18$ mag) and bright enough ($m_{F475W}<26.5$ mag) due to magnification to get ground-based spectroscopy in the rest-frame UV and optical bands.

\section{Completeness}
\label{sec:completeness}
The completeness correction factor $C(z,m,\mu)$ is the probability that a galaxy at redshift, $z$, with intrinsic apparent magnitude, $m$, and magnification, $\mu$, will be detected in our magnitude limited sample and also satisfy our color selection criteria. The completeness is affected by several factors: intrinsic luminosity, dust extinction, magnification, size and IGM opacity. In the subsequent subsections, we will describe how these quantities vary and how these variations are implemented in the completeness simulations.

\subsection{Simulating the Faint Galaxies}  
\label{subsec:sim_faint_galaxy}
We used BC03 models to generate a template spectrum by assuming a Chabrier initial mass function, constant star formation rate, 0.2 $Z_{\odot}$ metallicity, and an age of 100 Myr.  Both Salpeter \citep{sal55} and Chabrier \citep{cha03} initial mass functions roughly follow the same power law for stars with $M>1$M$_{\sun}$. However at smaller masses there are significant  differences.  This results in different stellar mass determinations, while having little effect on the ultraviolet spectral energy distributions.

We choose to use a somewhat younger, 100 Myr, stellar population \citep {hab12,hat13} than is typically used at these redshifts for star forming galaxies \citep[$\sim300$ Myr, e.g.][]{sha05,red08} as the target galaxies are typically smaller and have shorter dynamical timescales than their more massive counterparts. The SEDs don't change significantly for stellar ages between 100 Myr and 300 Myr \citep{lei99}, therefore our assumption for the age does not have a large effect on our main results (see Section \ref{sec:dust}). We will present the ages from SED fitting in a future paper (A. Dominguez et al., in preparation). The star formation histories are undoubtedly more complicated than the assumed constant rate, but the ultraviolet SED reacts slowly (on timescales of 50 Myr) to sudden changes \citep{lei99} and smooths out the effects of small timescale star formation events. 

Our galaxies have low stellar masses ($7< \log \left (M / M_{\sun}\right) < 9$; A. Dominguez et al., in preparation). Applying the mass-metallicity relation at these redshifts \citep{erb06, fyn08, bel13,yua13} to these galaxies gives a lower value than is typically assumed in similar studies which are based on brighter galaxies (0.2 vs 1 $Z_{\sun}$).

The Calzetti attenuation curve \citep{cal00} was used for the dust extinction by assuming a Gaussian distribution for $E(B-V)$ centered at 0.14 mag \citep{ste99,hat13} with a standard deviation of 0.1.

We also estimated the completeness for the typical stellar population assumptions (300Myr and 1Z$_{\sun}$)  and again a Gaussian distribution for $E(B-V)$ centered at a  value lower than 0.14 mag (0.05 mag, see Section \ref{sec:dust}) to see if varying the input stellar population parameters changes the results significantly. Our completeness determinations are robust against these initial considerations.

\subsection{Monte Carlo Simulation : IGM Opacity}
Ultraviolet (both Lyman emission line and continuum) photons are absorbed by neutral hydrogen clouds located along the line-of-sight to any galaxy. These intervening absorbers are classified in three groups based on their hydrogen column densities : Lyman-$\alpha$ forest ($10^{12}$ cm$ ^{-2}< N_{H}< 10^{17.5}$ cm$^{-2}$), Lyman limit systems (LLSs, $10^{17.5}$ cm$^{-2}< N_{H}< 10^{20.3}$ cm$^{-2}$) and damped Lyman-$\alpha$ systems (DLAs, $N_{H} > 10^{20.3}$ cm$^{-2}$). The column density distribution of these absorption clouds is given as

\begin{equation}
\frac{df}{dN_{H}}\propto N_{H}^{-k}
\end{equation}
and their number density distribution (number per unit redshift) changes as a power law with redshift.
\begin{equation}
N(z)=N_{0}(1+z)^{\gamma}
\label{equ:column_z}
\end{equation}
Where $N_{0}$ is the number of absorbers per unit redshift at present time (z = 0). The values we have adopted (from the literature) for k, N$_{0}$ and $\gamma$ are given in Table \ref {tab:IGM} \citep{jan06,rao06,rib11,ome12}

The IGM opacity at each line-of-sight is computed by running a Monte Carlo simulation. A complete description of this simulation is presented in \citet{sia08a}. Here we briefly summarize how the IGM absorption code works. 

We randomly vary the number of absorbers along the line of sight to each galaxy by sampling from a Poisson distribution with the expectation value $<N>$.
\begin{equation}
< N > =\int_{0}^{z_{\mathrm{galaxy}}}N(z)dz
\label{equ:expect_n}
\end{equation}
We then select a column density and redshift for the absorber from the distributions in Equations \ref{equ:column_z} and \ref{equ:expect_n}, respectively. We determine the Voigt profile with doppler parameters given in Table \ref {tab:IGM}, for the first twenty Lyman lines for each absorber. Finally, the IGM transmission for non-ionizing UV wavelengths at each line of sight is derived by adding the optical depths of randomly selected absorbers. We also incorporate the opacity to Lyman continuum photons using a photo-ionization cross section of $\sigma=6.3\times10^{-18}$ cm$^2$ and decreasing as $\lambda^{3}$ for $\lambda < 912$ \AA. 
We generated 1000 lines-of-sight at each redshift in steps of $\Delta z=0.1$ over the required redshift range for the completeness simulation ($1<z<3$, see Section \ref{sebsec:incomp_cor}).

\begin{deluxetable*}{llccccc}
\tablecaption{}
\tablewidth{0pt}
\tablehead{\colhead{Name} & \colhead{log(N$_{H}$)} & \colhead{k\tablenotemark{a}} & \colhead{N$_{0}$\tablenotemark{b}} & \colhead{$\gamma$\tablenotemark{b}} & \colhead{b (km s$^{-1}$)\tablenotemark{c}} & \colhead{redshift\tablenotemark{b}}  }
\startdata
Lyman-$\alpha$ forest & 12 - 14 & 1.67& 50.12 & 1.18 & 30 & $1.9 \leq z$ \\
 & & & 62.52 & 0.78 & 30 & $ z < 1.9$\\
 & 14 - 17.5 & 1.67 & 6.02 & 2.47 & 30 & $1.9 \leq z$\\
 & & & 16.98 & 1.66 & 30 & $ 0.7 \leq z < 1.9$\\
 & & & 35.4 & 0.13 & 30 &  $ z < 0.7$\\
 & & & & &\\
 Lyman limit system (LLS) & 17.5 - 20.3 & 1.07 & 0.17 & 1.33 & 70 & $ z < 2.6$ \\
 & & & & &\\
 Damped Lyman-$\alpha$ system (DLA)& 20.3 - 21.5 & 1.71 & 0.04 & 1.27 & 70 & $ z < 5  $\\
 & 21.5 - 22 & 11.1 & 0.04 & 1.27 & 70 & $z < 5 $\\
  
\enddata
\tablenotetext{a}{The values are taken from \citet{ome12}}
\tablenotetext{b}{The column density distribution parameters $N_{0}$ and $\gamma$ are from \citet{jan06,rib11,rao06} for Lyman-$\alpha$ forests, LLSs and DLAs , respectively.}
\tablenotetext{c}{The doppler parameter values are from \citet{kim97} and \citet{mol90}}

\label{tab:IGM}
\end{deluxetable*}

\begin{figure}
\epsscale{2.0}
\includegraphics[trim=0cm 0cm 0cm 0cm,clip=true,width=\columnwidth]{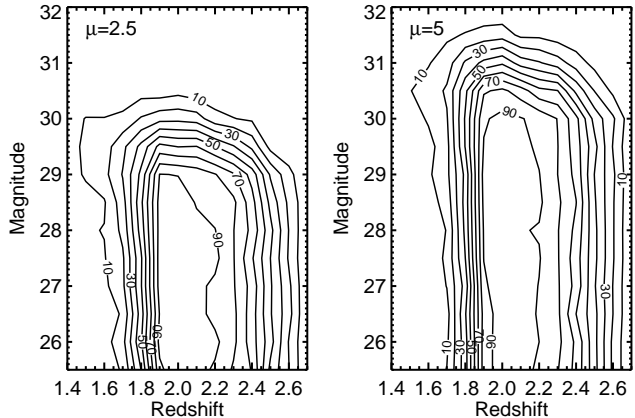}
\caption{The completeness contours as a function of intrinsic F475W apparent magnitude and redshifts. The contours are plotted for two values of magnification, $\mu$=2.5 mags (left, the typical magnification) and $\mu=5$ mags (right, an extreme magnification in our sample). }
\label{fig:comp}
\end{figure}

\subsection{Simulation: Incompleteness Correction }
\label{sebsec:incomp_cor}
We created a 3-D grid to compute the completeness as a function of redshift $z$, magnitude $m$ and magnification $\mu$. For each point in this 3-D space, the SED generated by BC03 was redshifted, magnified and then attenuated by the IGM for 300 randomly selected line-of-sights. At each line-of-sight, the dust attenuation is sampled randomly from the Gaussian distribution mentioned above.

As was mentioned before (see Section \ref{sec:color selection}), the primary contaminants in our sample are similar galaxies (LBGs) at slightly higher or lower redshifts. The photometric uncertainties can scatter the galaxies both into and out of the selection region. Adding these to the completeness simulation effectively broadens the redshift distribution of our sample. The photometric uncertainties are assumed to have a Gaussian distribution with $\sigma \propto \sqrt {\mathrm {area}}$ (assuming a constant instrumental noise over the object's area). We assume a normal distribution for the galaxy sizes centered at 0.7 kpc \citep{law12} with $\sigma$=0.2 kpc. At each line-of-sight, the randomly selected area is magnified by the magnification factor.

There is another important factor which can affect our completeness simulation. Charge transfer inefficiency (CTI) causes some signal to be lost during the readout. The CTI depends on the flux density of the object, sky background, the distance on the detector between the object and the readout amplifier and the time of the observation \citep[e.g. as the CTI worsens with time,][]{bag12,tep13}. This problem is more significant for images with very low background similar to our images in the F275W band \citep[$\sim0.5\ e^{-}$,][]{tep13}. Therefore CTI associated charge loses would make some galaxies fainter in the F275W band and push them inside the color selection region. We consider this issue in our completeness computations based on the analysis done in  \citet{noe12} . The CTI is estimated as follows:

\begin{equation}
\mathrm{m}_{(\mathrm{corr})}=\mathrm{m}_{(\mathrm{uncorr})}-\mathrm{S}\frac{\mathrm{Y}}{2048}
\end{equation}
where $Y$ is the distance in rows between the simulated galaxy and readout amplifier. For each line-of-sight, Y is randomly selected from a uniform distribution. S is a 2nd degree polynomial function of flux and observation date. More information related to the estimation of CTI is given in \citet{noe12}. The CTI corrections are small for most of the F275W-detected objects ($<0.1$ mag), therefore they do not have a significant effect on the completeness results. In Section \ref{sub:nondet}, we discuss about the possibility that CTI causes galaxies to be undetected in the F275W filter.

Therefore, at each point in the 3D grid ($z,m,\mu$), we simulate 300 galaxies at different lines of sight with different reddening, distance from readout amplifier and noise (size) selected from the related distributions mentioned above. The completeness was computed by counting the fraction of galaxies that satisfy the same selection criteria as the real observed objects (Equation \ref{equ:color-selection})

The completeness contours are given in Figure \ref{fig:comp}. The contours are plotted for two different values of magnification, $\mu=2.5$ and $\mu=5$ in units of magnitude which are the average and maximum predicted magnifications by the mass model for the LBG candidates in the sample. The figure indicates that the completeness as a function of absolute magnitude is obviously dependent upon the magnification. The color selection criteria represented in Section \ref{sec:color selection} is more than 90\% complete between $1.9<z<2.1$. 

As a test of the completeness simulation, we compare to the completeness of our selection of LBGs from a parent sample of galaxies with spectroscopic redshifts. The spectroscopic redshifts are discussed in J. Richard et al. 2013 (in preparation) and D. Stark et al. (in preparation). These spectroscopic campaigns have targeted primarily galaxies with high magnification (arcs) so many of the galaxies are intrinsically faint. In the core of the lensing cluster, we have targeted multiply imaged systems. Because the multiple systems are important for modeling the mass distribution of the cluster, they were targeted down to very faint magnitudes ($m_{475}\sim27.0$). Therefore the final spectroscopic sample is less biased toward luminous galaxies than it would be in a blank field without magnification. In the outskirts of the cluster, where the magnification is lower, spectra were obtained for a variety of galaxies.

\begin{figure}
\epsscale{2.0}
\includegraphics[trim=0cm 0cm 0cm 0cm,clip=true,width=\columnwidth] {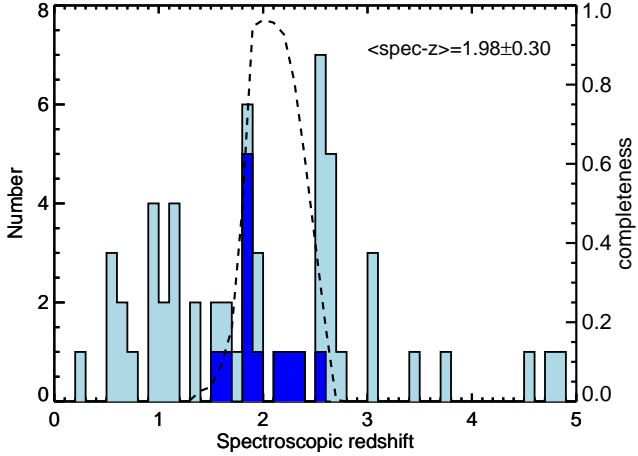}
\caption{The spectroscopic redshift distribution. The redshift histogram for galaxies with spectroscopic redshift is shown in light blue. The dark blue histogram shows the spectroscopic redshift distribution of 12 galaxies selected as $z\sim2$ LBGs in our image. We recovered 75\% (9 of 12) galaxies in the range $1.8<z<2.4$ where we should have high selection completeness. A few over densities are evident at $z\sim1.83$ and $z\sim2.54$. The right-hand axis shows the completeness values from the simulation. The dashed line shows the simulated completeness distribution over the redshift for a galaxy with typical intrinsic apparent magnitude ($m_{F475W}$=27.5) and magnification ($\mu=2.5$ mag).}
 
 {( A color version of this figure is available in the online journal)}
\label{fig:spec-z}
\end{figure}

Figure \ref{fig:spec-z} shows the redshift distribution of all 57 objects with spectroscopic redshifts in the field together with the subsample selected as LBGs. A few over densities are seen at z=1.83 and z=2.54. The dashed line is the completeness distribution for a galaxy with typical intrinsic apparent magnitude ($m_{F475W} = 27.5$) and magnification ($\mu = 2.5$ mag).  We detected 75\% of the galaxies in the redshift range, ($1.8 < z <2.4$), consistent with our completeness calculations. The simulated completeness values are given at right-hand axis. The tail in the dashed line seen at lower redshifts shows the contamination from lower redshift galaxies (see Section  \ref{sec:color selection}). This effect slightly broadens the redshift distribution of our sample, and is accounted for in the completeness simulation.

We see that the observed completeness, ratio of these two spectroscopic redshift histograms at each bin, is in good agreement with the corresponding simulated completeness value (dashed line). Predicting a high value of simulated completeness in the redshift bin $1.9<z<2$, we might expect to have more than one LBG out of three at this redshift interval. One of the objects was not selected because it is not a 5$\sigma$ detection in F336W.

\subsection{Redshift Distribution}
\label{subsec:redshift}

In order to determine luminosities and magnifications, we need to know the mean redshift of galaxies in our sample (and its dispersion). From our completeness function (Figure \ref{fig:comp}), if we assume that there is no strong evolution of number density with redshift, we would expect the average redshift of the sample to be between $2.0 \lesssim z \lesssim 2.1$. Given the unknown number density evolution and the possibility of structure along the line-of-sight, we determine the average redshift of the sample from both the spectroscopic and photometric redshifts.

From the 12 galaxies with spectroscopic redshift, we obtain an average redshift, $<z>=1.98$ with a dispersion of $\sigma_z=0.30$, in agreement with the completeness simulations. 

\begin{figure}
\epsscale{2.0}
\includegraphics[trim=0cm 0cm 1cm 0cm,clip=true,width=\columnwidth] {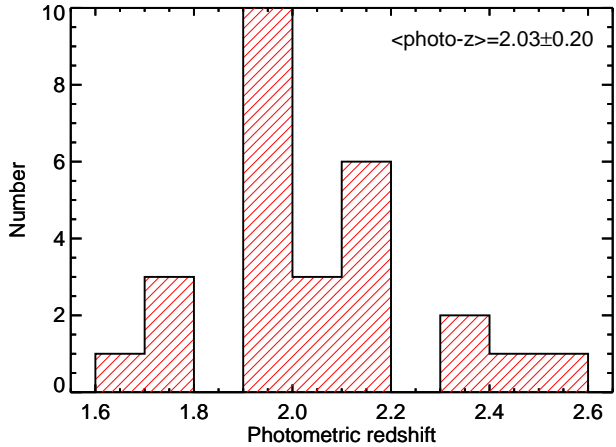}
\caption{The photometric redshift distribution from SED fitting to 8 {\it HST} bands spanning UV, optical and near-IR wavelengths.  The photometric redshift uncertainty is $\Delta=|(z_{phot}-z_{spec})|/(1+z_{spec})=0.02$ when comparing to the six galaxies in the sample with spectroscopic redshifts.}
 \label{fig:photo_z}
\end{figure}

To obtain a larger sample of redshifts, we calculated photometric redshifts for a sample of 26 galaxies that lie within the WFC3/IR image footprint. These photometric redshifts should be more precise because of the addition of the near-IR data, which span the Balmer/4000\AA\ break at $z\sim2$. We used the EAZY code \citep{bram08} to determine photometric redshifts of our eight {\it HST} band (UV,optical and IR) catalogs. We compare the photometric and spectroscopic redshifts (6 in this sample) to determine the fractional redshift error, $\Delta=|(z_{phot}-z_{spec})|\ /\ (1+z_{spec})=0.02$. The dispersion of the photometric versus the spectroscopic redshifts shows that there is no bias in our photometric redshift estimates. Figure \ref{fig:photo_z} shows the estimated photometric redshift distribution of the 26 candidates that have near-IR photometry. The mean of the photometric redshift distribution is $z=2.03$ with a dispersion of $\sigma_z=0.20$, in agreement with what we expect from the completeness simulation.

Given the average and dispersion of our spectroscopic and photometric redshifts, we assume an average redshift $< z >=2.00$ with a dispersion of $\sigma_z=0.25$ for LBGs without spectroscopic redshifts.  For comparison, the studies of \citet{oes10a} and \citet{hat10}, using similarly selected samples, assumed average redshifts of 1.9 and 2.1, respectively.

\subsection{F275W Non-detections and CTI}
\label{sub:nondet}
Our completeness simulations account for the small corrections to F275W magnitudes from CTI.  However, one concern is that very faint F275W fluxes near the detection limit will be lost completely due to CTI.  Of the final sample of 58 sources, 11 are undetected in F275W.  Most of these 11 galaxies are bright in F336W and would have bright F275W magnitudes if they were blue enough to lie outside of our selection window.  Therefore, CTI can not be responsible for the non-detection in F275W.  However, there are five galaxies with $F336W>27.0$ mag, meaning that the F275W magnitude of an LBG would have to be fainter than $F275W>28.0$ mag to be selected as an LBG.  For these galaxies, CTI can cause a non-detection if the galaxy is far from the readout amplifier.  The galaxies are 422, 1104, 1141, 1604, and 1620 pixels from the read amplifier.  It is possible that a few of the galaxies that are far from the amplifier ($>1000$ pixels) may be in our sample because of CTI issues.  We note however that these galaxies span a range in intrinsic UV magnitudes ($-16.98<M_{1500}<15.22$ mag) where there are many galaxies per bin.  Therefore, even in the worst case scenario that all four of these galaxies are low-z interlopers, the CTI concerns will not significantly affect the conclusions of this paper.

We are using two orbits of our cycle-20 program to test the effects of CTI in our F275W image and will refine our selection in the future.

\section{Luminosity Function}
\label{sec:LF}
The ultraviolet luminosity function at rest-frame 1500 \AA\  is measured by using the spectroscopic redshifts for 12 out of 58 dropout candidates  and assuming a mean redshift of 2.0 for the rest of the objects. The absolute magnitudes are computed at 1500 \AA\ by using the apparent magnitude at F475W as below:
\begin{equation}
M_{1500}=m_{F475W}+\mu-5\text{log}(d_{L}/10\text{ pc})+2.5\text{log}(1+z)
\end{equation}
Where $\mu$ is the magnification in magnitudes predicted by the lens model. The luminosity distribution of galaxies can be parametrized by a Schechter function which has three parameters: faint end slope ($\alpha$), characteristic luminosity ($L^*$) and normalization coefficient ($\phi$*).

\begin{figure*}
\epsscale{2.0}
\includegraphics[width=18cm]{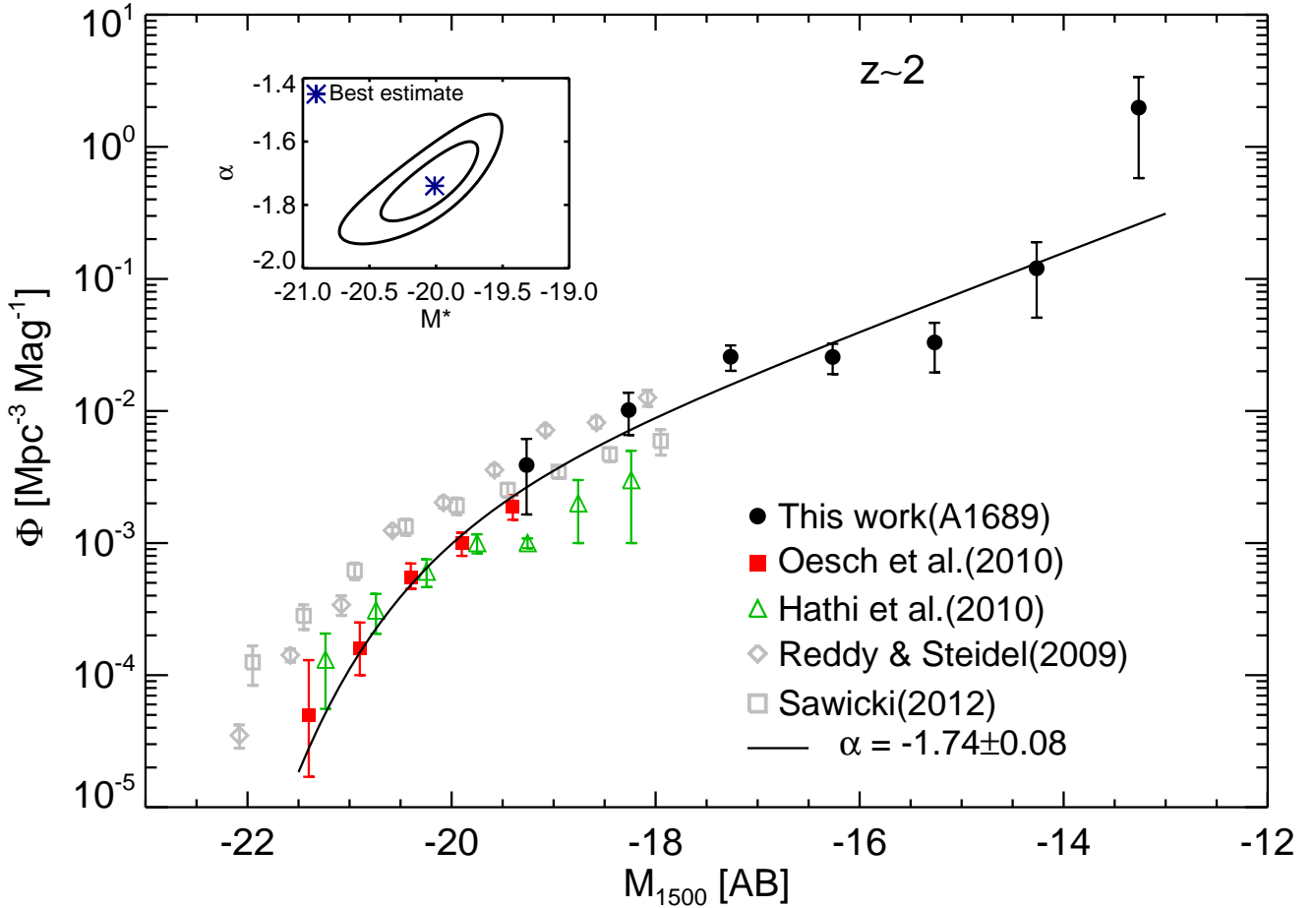}
\caption{The UV luminosity function at z$\sim$2. The black circles are our binned luminosity function values computed by using the Equation \ref{eqveff}. The red filled squares and the green open triangles are the LFs from  \citet{oes10a} and \citet{hat10}, respectively, which are selected in a similar manner to our sample and are at the same redshift ($z\sim2$). Grey open diamonds and grey open squares are the LFs at slightly higher redshift ($z\sim2.3$) from \citet{red09} and \citet{saw12}, respectively.
The black solid line is the best unbinned maximum likelihood fit to the whole sample of our data and \citet{oes10a}. The inset shows the 68\% (1$\sigma$) and 95\% (2$\sigma$) confidence contours of the Schechter parameters.}
{( A color version of this figure is available in the online journal)}
\label{fig:lf}
\end{figure*}

The accurate approach to fitting the luminosity function is the maximum likelihood method \citep{san79} using the individual galaxies and their associated effective volumes.
The main advantage of this approach is in using the unbinned data. Fits to binned data are imprecise as the choice of bin size and center effects the results. Furthermore, the completeness and the effective volume can change significantly from one side of the bin to the other. 

In addition to using our data in the maximum likelihood fitting process, we used the individual data points from \citet{oes10a} $z=1.9$ sample as well. Their sample includes the rare, bright galaxies which are not present in our small survey volume, but are necessary to constrain $L^{*}$.

The probability that a galaxy with absolute magnitude $M_{i}$ is detected in our magnitude limited sample is :
\begin{equation}
P(M_{i})=\frac{\phi(M_{i}) V_{\mathrm{eff}}(M_{i})}{\int_{-\infty}^{M_{\mathrm{limit}}} \phi(M) V_{\mathrm{eff}}(M) \mathrm{d} M}
\end{equation}

Where $\phi$ is the parametric luminosity function and $M_{limit}$ is the faintest intrinsic absolute magnitude in the sample. $V_{\mathrm{eff}}$ is the effective comoving volume in which a galaxy with magnitude $M$ can be found. The effective volume associated with each galaxy in our sample is derived by taking incompleteness into account,

\begin{equation}
V_{\mathrm{eff}}=\int_{\mu > \mu_{0}}\int_{0}^{\infty} \frac{dV_{com}}{dz d\Omega}C(z,m,\mu)\Omega(\mu) \mathrm{d}z \mathrm{d}\mu
\label{equ:v-eff}
\end{equation}
where V$_{\mathrm{com}}$ and $\Omega$ are the comoving volume and solid angle, respectively. $\Omega(\mu)$ is an area in the source plane that is magnified by a factor of $\mu$. Due to the distortion generated by gravitational lensing, the total effective area in the source plane is 0.37 arcmin$^{2}$ at $z\sim2$ which is  significantly smaller than the total area in the image plane. $\mu_{0}$ is the minimum magnification needed for detecting an object with magnitude $M$ relative to the magnitude limit. $C(z,m,\mu)$ is the completeness computed in  Section \ref{sec:completeness}.

The best fit to the luminosity function was found by maximizing the joint likelihood $\mathcal{L}$, which is the product of individual probabilities taken at all (unbinned) data points of the sample.
 
 \begin{equation}
 \label{eq5}
 \mathcal{L}=\prod_{i=1}^{N}P(M_{i})
 \end{equation}
 where $N$ is the total number of objects in both samples (Abell 1689 and \citet{oes10a}). In this method the normalization coefficient is cancelled out so it must be estimated separately from the number counts.
 \begin{equation}
 \phi^{*}=\frac{N}{\int_{M_{1}}^{M_{2}} \phi(M) V_{\mathrm{eff}}(M) \mathrm{d}M}
 \end{equation}
 Where $M_{1}$ and $M_{2}$ represent the brightest and faintest candidates in the sample, respectively.

There are three sources of uncertainty in the determination of the intrinsic absolute magnitudes for {\it our} sample. First, we do not know the redshift of most galaxies (except those with spectroscopic redshifts, Section \ref{sebsec:incomp_cor}), so the conversion from apparent to absolute magnitude is uncertain. 
We used the completeness computations given in Section \ref{sebsec:incomp_cor} (Figure \ref{fig:comp}) to estimate the standard deviation of the redshift distribution, $\Delta z=0.25$, which is consistent with the photometric and spectroscopic redshift distributions. Using this $\Delta z = 0.25$, we derive an absolute magnitude uncertainty of $\sigma_z =0.24$ magnitudes. Second, again due to the unknown source redshift, the magnification estimate from the mass model is uncertain. Using the same redshift uncertainty, $\Delta z = 0.25$, we find a $\sigma_{\mu}=0.10$ magnitudes. The third uncertainty is from the Abell 1689 lens model, $\sigma_{\mathrm{model}}$. \citet{lim07} has used a parametric method to describe the mass distribution of Abell 1689. In this parametric approach, they utilize the observational constraints (multiply imaged systems) to optimize the parameters (minimum $\chi^{2}$) of the lens model \citep[see][Equation 3]{lim07}. We computed the lens model uncertainty by producing a series of models through sampling the parameter space around a similar $\chi^{2}$ as the best model. We used a Bayesian Markov Chain Monte Carlo sampler in Lenstool \citep{jul07}. The mean and the median values of this uncertainty are 0.08 and 0.04 mags, respectively.

The total absolute magnitude uncertainty, $\sigma$, for the objects with spectroscopic redshift is simply $\sigma_{\mathrm{model}}$. Because the effect of luminosity distance and magnification when varying the redshift are correlated, they change the absolute magnitude in the same way. Therefore, $\sigma$ for the rest of the objects is computed by summing the first  two uncertainties, $\sigma_z$ and $\sigma_{\mu}$, and then adding the lensing model uncertainty in quadrature. This uncertainty in the absolute magnitude measurement is incorporated into the analysis by marginalizing over the probability distribution of each object's magnitude. The conditional probability that a galaxy with magnitude $M_{i}$ is in the sample, given the total uncertainty ($\sigma_{i}$) can be estimated through marginalization,

\begin{equation}
P(M_{i}|\alpha,\sigma_{i})=\int_{-\infty}^{\infty} P(M_{i}|\mu,\alpha,\sigma_{i}) P(\mu|\sigma_{i})\mathrm{d}\mu
\end{equation}
where $P(M_{i}\mid\mu,\alpha,\sigma_{i})$ is the Schechter function at magnitude $M_{i}$. The probability of each magnification measurement is given by a normal distribution,
\begin{equation}
P(\mu|\sigma_{i})=\frac{1}{\sqrt{2\pi} \sigma_{i}} e^{\frac{-(\mu-\mu_{i})^{2}}{2\sigma_{i}^{2}}}
\end{equation}
where the mean value, $\mu_{i}$, is the magnification assigned to each candidate by using the ratio of source and image plane luminosities from the Abell 1689 mass model \citep{lim07} and assuming an average redshift $z=2$ for all of the objects. 

 Figure \ref{fig:lf} shows the best-fit luminosity function derived from the maximum likelihood method along with previous determinations \citep{red09,hat10,oes10a,saw12}. The inset shows the 68\% and 95\% likelihood contours of the Schechter parameters. In order to display our data better (black filled circles in Figure \ref{fig:lf}), we defined the absolute magnitude bins and then calculated the luminosity function over these bins as below:
 \begin{equation}
 \label{eqveff}
\phi(M_{i})dM_{i}=\frac{N_{i}}{V_{\mathrm{eff}}(M_{i})}
\end{equation}
Where $N_{i}$ is the number of galaxies in magnitude bin i and $V_{\mathrm{eff}}$ is the effective comoving volume computed in Equation \ref{equ:v-eff}. We emphasize that we fit to individual points and not the binned data. 
 
We examined the convergence of the completeness simulation by splitting the simulation in half (150 line-of-sights each).  We then re-ran the simulation code and re-fit the luminosity function parameters using the maximum likelihood approach. The difference between the new estimates of each Schechter parameter and what we found before, is less than 4\% of  the previously determined uncertainty of each parameter. We conclude that our simulation has converged for the adopted number of line-of-sights (300, see Section \ref{sec:completeness}).

\begin{deluxetable*}{ccccc}
\tablecaption{UV Luminosity Function Parameters}
\tablewidth{0pt}
\tablehead{\colhead{Method} & \colhead{z} & \colhead{$\alpha$} & \colhead{M$^{*}$} & \colhead{$\mathrm{log}_{10}\phi$$^{*}$(Mpc$^{-3}$ mag$^{-1}$)}}
\startdata
Maximum Likelihood\tablenotemark{a} & 2.0 & -1.74$\pm$0.08 & -20.01$\pm$0.24 & -2.54$\pm$0.15
\enddata
\tablenotetext{a}{Maximum likelihood fit to the whole sample including both this work (Abell 1689) and \citet{oes10a}}
\label{tab:LF_param}
\end{deluxetable*}

\subsection{Cosmic Variance}
\label{subsec:cv} 
The cosmic variance uncertainty $\sigma_{\mathrm{CV}}$ in the galaxy number counts can be estimated through the effective volume of the survey, the survey geometry, and an estimate of the typical clustering bias of the discovered sources. In what follows, we compute the cosmic variance uncertainty for the lensed field.

The effective volume of our survey has been calculated using the methods described in Section \ref{sec:LF}. We use these effective volumes and the selection function of the survey with redshift to determine the root-mean-squared (RMS) density fluctuations $\sigma_{\rho}$ expected in our survey volume given its pencil beam geometry, following the methodolgy of \citet{rob10}. We find these density fluctuations to be $\sigma_{\rho}\approx0.1$, which is determined largely by the line-of-sight extent of the pencil beam survey (the comoving radial distance over the redshift range $1.75\le z \le 2.35$ where our selection is efficient) and the linear growth factor $D(z\sim2)\approx0.4$ \citep{rob13}.

To determine the cosmic variance uncertainty in the galaxy counts, we perform a simple abundance matching calculation \citep[e.g.,][]{con06,con09} assigning galaxies in our survey approximate
halo masses and clustering bias based on their volume abundances.  For galaxies in our survey, the estimated bias is $b \sim 1.2-2.6$, providing a cosmic variance uncertainty of $\sigma_{\mathrm{CV}}\approx0.12-0.25$ \citep[e.g.,][]{rob10}, comparable
to our fractional Poisson uncertainty $1/\sqrt{N}\approx 0.13$. We therefore expect that cosmic variance does not strongly influence the luminosity function results.  Further, since cosmic variance instills a covariance in the galaxy number counts as a function of luminosity \citep[see, e.g.,][]{rob10}, if our survey probes either an over- or under-dense region compared to the cosmic mean the covariance in the counts should
have little effect on the intrinsic shape of the luminosity function (especially at faint magnitudes where the galaxies are nearly unbiased tracers of the dark matter).  Our faint-end slope determination is therefore expected to be robust against systematic considerations owing to cosmic variance uncertainties.

\section{UV Spectral Slope}
\label{sec:dust} 
The ultraviolet continuum of galaxies can be approximated as a power law, $f_{\lambda}\propto\lambda^{\beta}$ \citep{cal94}. The UV spectral slope, $\beta$, of each galaxy in our sample was estimated by making fake power law spectra over a wide range of $\beta$ values and multiplying these spectra with the filter curves. We then have a one-to-one map of the observed color to the spectral slope. We use the F475W and F625W filters to measure the UV spectral slope as they correspond to rest-frame wavelengths of $\sim1580$ \AA\ and $\sim2080$ \AA\ respectively, at $z\sim2$. The uncertainty of the $\beta$ estimate for each individual object was derived by using the photometric uncertainties in both the F475W and F625W filters. 

The $E(B-V)$ values, which have a one to one relation with UV spectral slopes, are obtained based on the comparison of observed UV colors with the colors predicted from the stellar population synthesis model (BC03) and reddened with a Calzetti attenuation curve \citep{cal00}. We used the realistic assumptions for the age (100 Myr) and the metallicity ($0.2\ Z_{\sun}$) of these compact faint galaxies (see Section \ref{subsec:sim_faint_galaxy} ). In order to compare the importance of the age and metallicity in the color excess measurements, we also estimated the $E(B-V)$ values considering the more typical assumptions for these two quantities ($1\ Z_{\sun}$ \& 300 Myr). We present the color excess $E(B-V)$ distributions for three sets of age and metallicity assumptions, [($0.2\ Z_{\sun}$ \& 100 Myr), ($0.2\ Z_{\sun}$ \& 300 Myr), ($1\ Z_{\sun}$ \& 100 Myr)] in Figure \ref{fig:ebv_hist}. The black histogram shows the distribution resulting from the realistic assumptions for the age and metallicity ($Z=0.2\ Z_{\sun}$ and age $=100$ Myr). We see that varying the assumed metallicity (right panel, blue distribution) dominates the effect on the determined reddening distribution, whereas changing the assumed starburst age has little effect (left panel, red distribution). Therefore, assuming a value of $1\ Z_{\sun}$ would underestimate the dust content of these faint galaxies.

\begin{figure*}
\epsscale{2.0}
\includegraphics[trim=0cm 0cm 0cm 0cm,clip=true,width=18cm] {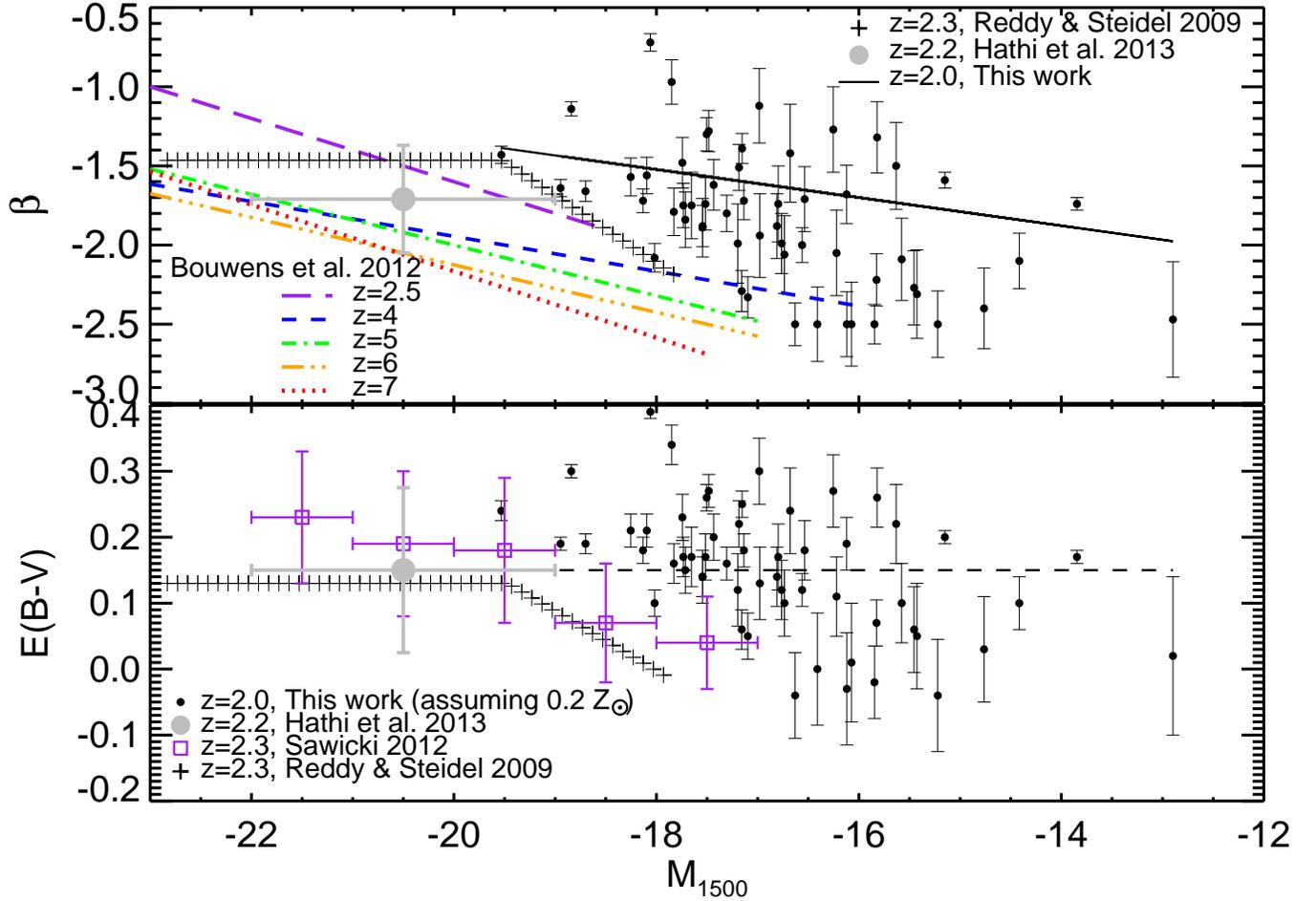}
\caption{Upper panel: UV spectral slope as a function of absolute UV magnitude at 1500 \AA. The black solid line is the best linear fit to our individual data. The line made of plus symbols is from \citet{red09} . The purple long dashed, blue dashed, green dot-dash, orange dot-dot-dash and red dotted lines are the best linear fit for redshifts 2.5, 4, 5, 6 and 7 respectively, from \citet{bou12b}. The grey point shows the average and dispersion of values measured by \citet{hat13} at $z\sim2$. Lower panel: E(B-V) as a function of absolute UV magnitude at 1500 \AA\ . Our reddening values (black filled circles) are computed assuming $0.2\ Z_{\sun}$ but the other E(B-V) values have been found assuming $1\ Z_{\sun}$. The purple open squares are from \citet{saw12}. The rest of the symbols are the same as those in the upper panel. The dashed line is an average of our E(B-V) values and {\it not} a fit to our data.}
{( A color version of this figure is available in the online journal)}
\label{fig:beta}
\end{figure*}

In the upper panel of Figure \ref{fig:beta} we plot the estimated UV slopes versus absolute UV magnitude at 1500 \AA\ ($M_{1500}$). The UV spectral slope, $\beta$, correlates with the UV magnitude, $M_{1500}$, such that the less luminous galaxies are bluer. The same trend has been identified in other works at higher redshifts and higher luminosities \citep{meu99,lab07,ove08,wil11,bou12b}. Fitting the estimated $\beta$ values versus UV luminosities, we find $\beta$=(-0.09$\pm$0.01)$M_{1500}$+(-3.1$\pm$0.1) for our z$\sim$2 dropout sample. We note that we removed one galaxy from our sample when fitting the reddening distribution as this galaxy had an anomalously high F625W flux. Therefore the derived reddening was quite high, even though the photometry in other bands suggested a blue spectrum.

\begin{figure}
\epsscale{2.0}
\includegraphics[trim=0cm 0cm 0cm 0cm,clip=true,width=\columnwidth] {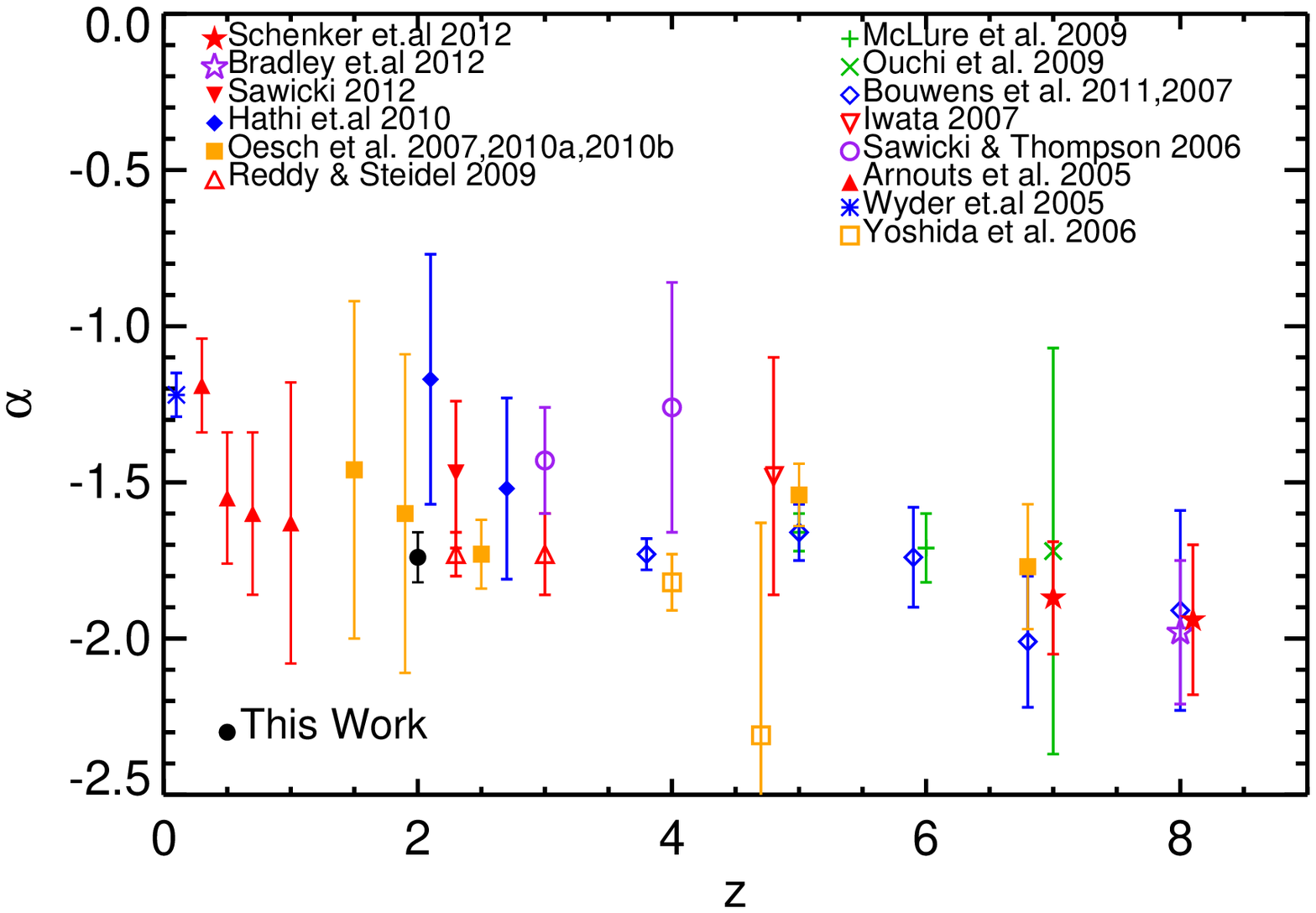}
\caption{The evolution of the faint-end slope with redshift. The black filled circle shows our measurement of $\alpha$. The plot also includes other determinations from the literature \citep{arn05,wyd05,saw06,yos06,bou07,iwa07,oes07,mcl09,ouc09,red09,oes10a,oes10b,hat10,bou11,bra12,saw12,sch12}. The data points at z=8 are all at the same redshift but have been offset slightly for clarity. } 
{( A color version of this figure is available in the online journal)}
\label{fig:alpha-z}
\end{figure}

\section{DISCUSSION}
\label{sec:discussion}
We have extended the UV luminosity function at $z\sim2$ to the faintest magnitude limit ever obtained ($M_{1500}\sim-13$ mag), allowing us to put a strong constraint on the ultraviolet luminosity density at the epoch of peak star formation. In the next few subsections, we will discuss the contribution of faint galaxies to the ultraviolet luminosity and star formation densities at $z\sim2$. We will also discuss the evolution of the faint-end slope of the luminosity function and dust extinction derived from the UV spectral slope, in comparison with other results from the literature.

\subsection{Comparison with Other Space Density Measurements at $z\sim2$ }

In Figure \ref{fig:lf}, we plot our binned luminosity function with those of \citet{red09, oes10a, hat10, saw12}. We prefer to compare our luminosity function with those of \citet{oes10a} and \citet{hat10}, as the galaxies are selected via a Lyman break in the same F275W filter and will therefore have similar redshift distributions.  The \citet{red09} and \citet{saw12} samples are selected with optical data alone and are at a slightly higher average redshift ($z\sim2.3$).  Given the significant evolution in the luminosity function at these redshifts \citep{oes10a}, a direct comparison with the \citet{red09} and \citet{saw12} data is not ideal but the data are plotted for reference.  

If we assume that the faint-end of the luminosity function is truly a power-law, then the fit to that power law suggest an excess of moderately bright galaxies ($M_{1500}\sim-19$ mag) in our sample compared to the measurements of \citet{oes10a} and \citet{hat10}, though the space densities are consistent within the $1\sigma$ uncertainties. We note that both of these measurements \citep{oes10a,hat10} are from the same data, a 50 arcmin$^2$ area in the GOODS-South region as part of the {\it HST} Wide-Field Camera 3 Early Release Science (ERS) data \citep{win11}. It is important to note that both the ERS bright galaxy sample and our sample may suffer from sample variance, because they are from single, small area fields. Though the ERS UV data covers eight times the area of our survey, they are also probing more massive galaxies, which are more highly biased and subject to sample variance. Therefore surveys in additional fields are needed to address the sample variance at both the bright and faint end of the LF.

\subsection{Evolution of the Faint-End Slope}
Using strong gravitational lensing, we have extended the measurement of the space density of $z\sim2$ galaxies two orders of magnitude fainter than previous studies, allowing a more precise measure of the faint-end slope of the UV luminosity function. In Figure \ref{fig:alpha-z}, we compare our measurement of the faint-end slope with results at other redshifts. A general implication of this plot is that $\alpha$ is steeper at high redshifts than at lower redshifts. The evolution of the faint-end slope is slow between $2<z<7$, however it has evolved significantly between $z=2.3$ \citep[$\alpha\sim-1.73$][]{red09} and the present \citep[$\alpha\sim-1.2$][]{wyd05}.  Our faint-end slope measurement, $\alpha=-1.74$, is nearly consistent with the estimates at higher redshifts ($2.5<z<4$) within the error bars. 
Additional sight lines and deeper selection of UV galaxies at $z\sim1$ will help us constrain the evolution of the faint end slope over the last 10 Gyr.

\subsection{Evolution of Dust in Faint Star-Forming Galaxies}
\label{sub:redder_galaxies}

In the upper panel of Figure \ref{fig:beta} we show the $\beta$ vs. $M_{1500}$ relation of our sample (black line) and similarly selected samples at higher redshift \citep{red09, hat13, bou12b}.  We see the same trend that is seen at higher redshifts in that fainter galaxies have bluer UV spectral slopes.  The measured slope of the trend, $d\beta$ / d$M_{1500} = -0.09 \pm 0.01$, is similar to the slopes measured at higher redshift, though the zeropoint is offset such that galaxies of the same UV luminosity are redder at later epochs. This is consistent with the trend seen from $2.5 < z < 7$ \citep{bou12b}.  The increase in $\beta$ at the same absolute magnitude from $z\sim2.5$ to $z\sim2$ (time difference of $\sim 600$ Myr) is about 0.4, consistent with the increase in beta from $z \sim 4$ to $z \sim 2.5$ (time difference of $\sim 1$ Gyr).

The luminosity-dependent UV spectral slope of \citet{red09} (derived from their reddening estimates) is also plotted in Figure \ref{fig:beta} (plus symbols). Our best fit comes very close to their constant value of $\beta$ at $M_{1500}\sim-19.5$ mag and consistent with the average value measured by \citet{hat13} at the same redshift. The difference is that \cite{red09} see bluer galaxies than our sample at $-19.5<M_{1500}<-18$ mag. In the faintest bin, $M_{1500}\sim-18$ mag, \citet{red09} state that the colors are consistent with no dust extinction, whereas we see significant extinction in galaxies at this absolute magnitude. 

We believe that there are two explanations for the redder average spectral slopes exhibited in our sample compared to the \citet{red09} sample. First, there is a 400 Myr time difference between the average redshift of the \citet{red09} sample ($z=2.3$) and our sample ($z=2$), and we know that galaxies of the same luminosity are getting redder with time \citep{bou12b}. Second, we are likely less biased against detecting red galaxies at these absolute magnitudes. Our Lyman break selection is more complete than selections like BM/BX \citep{ade04}, which purposely target galaxies that are blue in all filters. Also, our ultra-deep imaging along with the high magnification allows us to detect redder, fainter objects at higher $S/N$ in the bluest bands. 

\begin{figure}
\epsscale{2.0}
\includegraphics [trim=0cm 6cm 0cm 0cm,clip=true,width=\columnwidth] {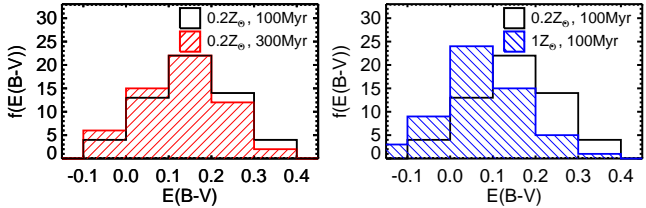}
\caption{The $E(B-V)$ distributions of the $z\sim2$ LBG sample. The $E(B-V)$ values are derived by comparing the LBG colors with the dust reddened colors predicted by stellar population models. The black distribution is computed by assuming the realistic values for the age (100 Myr) and metallicity (0.2 Z$_{\sun}$) of the stellar population models (see Section \ref{subsec:sim_faint_galaxy}). The red (left panel) and the blue (right panel) hatched histograms show the $E(B-V)$ values measured by assuming (0.2 Z$_{\sun}$ \& 300 Myr) and (1 Z$_{\sun}$ \& 100 Myr), respectively.  The assumed age of the galaxy (left) has little effect on the derived reddening values, whereas the assumed metallicity (right) has a very large effect.  If we were to assume solar metallicity, a large fraction of galaxies would have colors consistent with no extinction at all.}
\label{fig:ebv_hist}
\end{figure}

The measured reddening of our sample is strongly dependent on the assumed stellar population parameters, as they affect the intrinsic UV spectral slope. This is exhibited in Figure \ref{fig:ebv_hist}, where we show that changing the metallicity of the stellar population from $0.2\ Z_{\sun}$ to $1.0\ Z_{\sun}$ decreases the average color excess from $E(B-V)=0.15$ mag to $E(B-V)=0.08$ mag.  The assumed starburst age has very little effect (average $E(B-V)=0.15$ mag to $E(B-V)=0.13$ mag when increasing the age from 100 to 300 Myr).  

The lower value of metallicity we have used in this work, $Z=0.2\ Z_{\sun}$, is justified by the low stellar masses of our sample galaxies ($7<\log \left (M / M_{\sun} \right) <9 $, Dominguez et al., in preparation) and the metallicity measurements at high and low mass at these redshifts \citep{erb06, bel13}, assuming that stellar and gas-phase metallicities are similar. 

The ultimate goal in measuring the UV spectral slopes of these galaxies is to infer the extinction of the ultraviolet light in order to measure the intrinsic UV luminosities and star formation rates. In the lower panel of Figure \ref{fig:beta} we show the implied color excess, $E(B-V)$, given the intrinsic spectral slope of our fiducial model (constant star formation, age $= 100$ Myr, $Z=0.2\ Z_{\sun}$) and assuming a Calzetti attenuation curve.

Intriguingly, nearly every galaxy brighter than $M_{1500}<-15$ mag has significant reddening. The dashed line in the lower panel of Figure \ref{fig:beta} shows the average color excess of our sample , $<E(B-V)> = 0.15$ mag, which is similar to the value measured for much more luminous galaxies by \citet[][$E(B-V)=0.13$ mag, plus symbols]{red09},  \citet[][purple squares]{saw12}, and \citet[][$E(B-V)=0.15$ mag, grey filled circle]{hat13}, who assumed solar metallicity.

Thus, we come to the important conclusion that the trend of bluer UV spectral slopes at fainter absolute magnitudes is {\it not} necessarily due to decreasing dust reddening.  Rather, the dust reddening at faint magnitudes ($-18<M_{1500}<-15$ mag) is similar to the reddening in more luminous galaxies ($-21<M_{1500}<-18$ mag), and the bluer observed UV slopes are due to bluer {\it intrinsic} UV slopes because the stellar population is metal poor. Of course, the reddening likely depends on luminosity as well. In order to know the exact relation of average extinction with luminosity, we need a more accurate measure of the luminosity- (or mass) metallicity relation. Furthermore, this analysis has assumed a Calzetti attenuation curve.  There is some evidence that young galaxies may have steeper attenuation curves \citep[e.g. SMC,][]{sia08b,sia09,red10}.  Measurements of the infrared luminosities of these faint galaxies will help determine which attenuation curve is more appropriate.  

In the future, measurements of metallicities (with rest-frame optical spectroscopy) and infrared luminosities will help us better understand the extinction in these faint galaxies. Because of the high magnification, these galaxies comprise an ideal sample for follow-up.

\subsection{The Effect of Intracluster Dust }
In this study, all of the UV-dropouts are located behind a massive cluster so the light coming from these background galaxies can be affected by intracluster dust. Recent studies of SDSS clusters \citep{che07,bov08,mul08} have shown that there is a negligible amount of intracluster dust attenuation $E(B-V) < 8\times10^{-3}$ mag on scales smaller than 1 Mpc from the cluster center. We estimated the average intracluster dust reddening($A_{\lambda}$) in the UV and optical bands to see if it has any effect on our LBG selection or spectral slope estimates. 
\citet{bov08} measures the intracluster dust attenuation for a large sample of SDSS clusters. They calibrate the extinction curve presented in \citet{cha00}, by comparing the spectra of  galaxies that lie behind and adjacent to the SDSS clusters. We approximated the $A_{\lambda}$ values based on their calibrated extinction curve. Both the UV-dropout selection and  the UV spectral slope measurements are not significantly affected by intracluster reddening because the estimated color excesses are negligible ($A_{F275W}-A_{F336W}=0.01$ mag, $A_{F475W}-A_{F625W}=0.01$ mag).  

\subsection{UV Luminosity Density}
\label{sub:uv-density}

In the following discussion we prefer to compare to the samples of \citet{oes10a} and \citet{hat10} because the samples are selected with similar filters and are at a similar redshift.  If we integrate our luminosity function over all luminosities down to zero to find the entire UV luminosity density, we get the value of $\rho_{UV}=43.1\substack{+6.8 \\ -6.0}\times10^{25}$ erg s$^{-1}$ Hz$^{-1}$ Mpc$^{-3}$. A large fraction of $71\%$ of the total luminosity density at $z=2$ is from the luminosity range of our sample alone ($-19.76<M_{1500}<-12.76$ mag). The fraction is less than $10\%$ for galaxies in the absolute magnitude range of \citet{oes10a} which are brighter than our sample ($M_{1500}<-19.76$ mag). Our most luminous galaxy is about the faintest galaxy seen in the \citet{oes10a} sample, so there is very little overlap in luminosities. The faint galaxies in our sample account for seven times more UV luminosity density than the brighter galaxies from \citet{oes10a}. If we assume the luminosity function has the same slope down to zero luminosity, integrating from our faintest bin down to zero only increases the UV luminosity density by $20\%$. All of these values are given in Table \ref{tab:UV_lum}. We note that extending the luminosity range to much larger luminosities adds a negligible amount to the UV luminosity density. This demonstrates the power of cluster lensing to quickly uncover the primary sources of star formation at these epochs.

\subsection{Star Formation Rate Density}
The evolution of the star formation rate density (SFRD) has been an ongoing subject of research, especially at $1<z<3$ because star formation appears to have peaked at this epoch \citep[e.g.,][]{cal99,ste99,fer00,hop00,hop06,red09}. Because of the steep faint-end slope of the LF, much of the star formation occurs in faint galaxies. Furthermore, because of the significant extinction seen in our sample, there is even more star formation in our faint sample.  

Of course, because this population is not well studied, many of the assumptions typically applied in such studies may not apply to our sample. First, as mentioned previously, the metallicity is likely significantly lower than Solar, which results in significant ($\sim15$\%) differences in the conversion of UV luminosity density to SFRD.  Second, the starburst ages may be significantly younger than $10^8$ yr. For younger starbursts, the star formation rate is a function of both the UV luminosity and the population age. Indeed, the assumption of continuous star formation may not be accurate at all in these systems where supernovae are thought to be very effective at shutting down star formation on short time scales \citep[e.g. ][]{gov12}.  Third, because the typical attenuation curve in such systems has not been well measured, the dust corrections are still not well understood.  Given these caveats, we calculate below the best estimate of the SFRD from UV-selected galaxies at $z\sim2$.

We use the \citet{ken98} conversion of the UV luminosity density to SFR as below: 

\small
\begin{equation}
\mathrm{SFR} (\text{\ M}_{\sun}\text{ yr}^{-1})=1.4\times10^{-28}L_{\mathrm{UV}}\ (\text {erg s}^{-1} \text{Hz}^{-1})
\end{equation}
\normalsize

\noindent where a Salpeter initial mass function \citep{sal55} from 0.1-100 $M_{\sun}$ is assumed. Using the total UV luminosity density from Table \ref{tab:UV_lum}, we find a SFRD uncorrected for dust of 0.060 $M_{\sun}$ yr$^{-1}$ Mpc$^{-3}$.  To correct for extinction, we note that the average attenuation measured in our sample ($E(B-V)=0.15$ mag) is similar to the values measured at the bright end by \citet{red09} and \citet{hat13}.  Thus, we use this constant value to derive a factor of 4.17 correction for extinction (assuming a Calzetti attenuation curve) for galaxies of all luminosities. We therefore determine a SFRD of 0.252$\substack{+0.040 \\ -0.035}$ $M_{\sun}$ yr$^{-1}$ Mpc$^{-3}$ for UV-selected galaxies at $z\sim2$.  If we use a Kroupa \citep{kro01} or Chabrier \citep{cha03} IMF, this value needs to be divided by 1.7 or 1.8, respectively to account for the decreased number of low mass stars relative to the Salpeter IMF.  We convert to a Kroupa IMF (SFRD$_{\mathrm{Kroupa}}=0.148\substack{+0.023\\ -0.020}$ M$_{\sun}$ yr$^{-1}$ Mpc$^{-3}$) to compare to the value estimated by \citet[][0.122$\substack{+0.027 \\ -0.027}$ M$_{\sun}$ yr$^{-1}$ Mpc$^{-3}$]{red09}.  Our value is about 20\% higher but within their error bars. It is important to note however, that \citet{red09} use an average dust extinction correction factor of 1.91, less than half the correction that we use (4.17).  Thus, if they implemented the same constant extinction correction as this paper for all UV-selected galaxies, their estimate of the SFRD would more than double. This shows the importance of the dust correction estimate for the fainter galaxies.
 
\begin{deluxetable}{cc}
\tablecaption{UV Luminosity Density}
\tablewidth{0pt}
\tablehead{\colhead{Range\tablenotemark{a}} & \colhead{UV Luminosity Density\tablenotemark{b}}}
\startdata
[-21.68,-19.76]\tablenotemark{c}	      &    4.16$\substack{+0.46 \\ -0.49}$\\[0pt]
[-19.76,-12.76]                                              &    30.8$\substack{+1.9 \\ -3.1}$\\[0pt]
[-12.76,0.00]	 			               &     8.09$\substack{+4.4\\ -2.4}$ \\
\hline
[-21.68,0.00]					      &	   43.1$\substack{+6.8 \\ -6.0} $ 

\enddata
\tablenotetext{a}{Absolute magnitudes at 1500 \AA.}
\tablenotetext{b}{units of $\times10^{25}$ erg s$^{-1}$ Hz$^{-1}$ Mpc$^{-3}$}
\tablenotetext{c}{The \citet{oes10a} limit is slightly fainter, but we only integrate to our bright limit}
\label{tab:UV_lum}
\end{deluxetable}

\section{Summary}
\label{sec:result}
The sensitive ultraviolet imaging capability of the WFC3/UVIS camera allows us to study intermediate redshift ($1<z<3$) star-forming galaxies. We used the deepest near-UV images ever obtained with the {\it HST}/UVIS channel to identify ultra-faint star-forming galaxies located behind the massive cluster Abell 1689. We found 58 Lyman break galaxies at $z\sim2$ that are highly magnified due to strong gravitational lensing. The main conclusions of this work are summarized below:

\begin{enumerate}

\item The faint-end slope of the UV luminosity function is estimated to be $\alpha=-1.74\pm0.08$, consistent with previous determinations at $2.3<z<6$ \citep{bou07,red09}. 

\item The UV luminosity function shows no turnover down to very faint UV magnitudes ($M_{1500}\sim-13$ mag). This is particularly interesting because the bright sources do not provide sufficient ionizing photons to ionize the universe by $z\sim6$ \citep{rob13}. Galaxies of such low luminosities are required at $z>7$ to reionize the intergalactic hydrogen and produce the high Thompson scattering optical depth to the cosmic microwave background seen by the Wilkinson Microwave Anisotropy Probe \citep{kuh12, rob13}.  Indeed, these faint galaxies may contribute significantly to the ionizing background at moderate redshift \citep[$z\sim3$, ][]{nes12}.

However recent numerical simulations by \citet{kuh13} show a cutoff in the simulated UV LF at $M_{1500}$=-16, due to the suppression of the star formation in low metallicity faint galaxies. The  discrepancy between our results and \citet{kuh13} is only in our two faintest magnitude bins, where the number of objects is small.  Further LBG searches in more lensing clusters will provide a more robust test of this prediction.

\item The UV spectral slope, $\beta$, for these LBGs at $z\sim2$ is redder than higher redshift determinations at the same UV luminosities. The correlation between $\beta$ and the rest-frame UV magnitude implies higher dust extinction in more luminous galaxies, as is seen at higher redshifts.  We find evidence for significant dust extinction, averaging $E(B-V)\sim 0.15$ mag, in most star-forming galaxies with $-18<M_{1500}<-15$ mag, in contrast with previous measurements at these redshifts \citep{red09, saw12}.  Our finding assumes a Calzetti attenuation curve and $Z=0.2\ Z_{\sun}$ metallicity.  Both assumptions need to be confirmed with further studies.

\item We derive a total UV luminosity density of $4.31\substack{+0.68 \\ -0.60}\times10^{26}$ erg s$^{-1}$ Hz$^{-1}$ Mpc$^{-3}$ when integrating our luminosity function and extrapolating to zero luminosity. More than 70\% of the UV luminosity density originates from the galaxies in the luminosity range covered by our sample.  We estimate that no more than 20\% of the UV luminosity density originates from fainter galaxies than those in our sample.

\item Assuming a constant extinction ($E(B-V)=0.15$ mag, UV dust correction of 4.2) for galaxies of all luminosities, we estimate the global SFRD (of UV-selected galaxies) to be 0.148$\substack{+0.023 \\ -0.020}$  $M_{\sun}$ yr$^{-1}$ Mpc$^{-3}$ (Kroupa IMF). This number is dependent on many assumptions regarding the ages, metallicities and extinction curves for this faint population of galaxies.  Further investigations are required to accurately determine these properties of this new population.

\end{enumerate}
The authors wish to thank Pascal Oesch who kindly sent his data to us for our LF computations. We also thank the referee for useful suggestions, as well as Marcin Sawicki for his valuable comments.
This work is based on observations with the NASA/ESA Hubble Space Telescope, obtained at the Space Telescope Science Institute, which is operated by the Association of Universities for Research in Astronomy, Inc., under NASA contract NAS 5-26555.

{\it Facilities:} \facility{{\it HST} (WFC3,ACS)}

\bibliographystyle{apj}
\bibliography{citation.bib}

\begin{thebibliography}{113}
\expandafter\ifx\csname natexlab\endcsname\relax\def\natexlab#1{#1}\fi

\bibitem[{{Adelberger} {et~al.}(2004){Adelberger}, {Steidel}, {Shapley},
  {Hunt}, {Erb}, {Reddy}, \& {Pettini}}]{ade04}
{Adelberger}, K.~L., {Steidel}, C.~C., {Shapley}, A.~E., {et~al.} 2004, \apj,
  607, 226

\bibitem[{{Arnouts} {et~al.}(2005){Arnouts}, {Schiminovich}, {Ilbert},
  {Tresse}, {Milliard}, {Treyer}, {Bardelli}, {Budavari}, {Wyder}, {Zucca}, {Le
  F{\`e}vre}, {Martin}, {Vettolani}, {Adami}, {Arnaboldi}, {Barlow}, {Bianchi},
  {Bolzonella}, {Bottini}, {Byun}, {Cappi}, {Charlot}, {Contini}, {Donas},
  {Forster}, {Foucaud}, {Franzetti}, {Friedman}, {Garilli}, {Gavignaud},
  {Guzzo}, {Heckman}, {Hoopes}, {Iovino}, {Jelinsky}, {Le Brun}, {Lee},
  {Maccagni}, {Madore}, {Malina}, {Marano}, {Marinoni}, {McCracken}, {Mazure},
  {Meneux}, {Merighi}, {Morrissey}, {Neff}, {Paltani}, {Pell{\`o}}, {Picat},
  {Pollo}, {Pozzetti}, {Radovich}, {Rich}, {Scaramella}, {Scodeggio},
  {Seibert}, {Siegmund}, {Small}, {Szalay}, {Welsh}, {Xu}, {Zamorani}, \&
  {Zanichelli}}]{arn05}
{Arnouts}, S., {Schiminovich}, D., {Ilbert}, O., {et~al.} 2005, \apjl, 619, L43

\bibitem[{{Baggett} {et~al.}(2012){Baggett}, {Noeske}, {Anderson}, {MacKenty},
  \& {Petro}}]{bag12}
{Baggett}, S.~M., {Noeske}, K., {Anderson}, J., {MacKenty}, J.~W., \& {Petro},
  L. 2012, in Society of Photo-Optical Instrumentation Engineers (SPIE)
  Conference Series, Vol. 8453, Society of Photo-Optical Instrumentation
  Engineers (SPIE) Conference Series

\bibitem[{{Belli} {et~al.}(2013){Belli}, {Jones}, {Ellis}, \&
  {Richard}}]{bel13}
{Belli}, S., {Jones}, T., {Ellis}, R.~S., \& {Richard}, J. 2013, ArXiv e-prints

\bibitem[{{Bertin} \& {Arnouts}(1996)}]{ber96}
{Bertin}, E., \& {Arnouts}, S. 1996, \aaps, 117, 393

\bibitem[{{Bouwens} {et~al.}(2012{\natexlab{a}}){Bouwens}, {Bradley}, {Zitrin},
  {Coe}, {Franx}, {Zheng}, {Smit}, {Host}, {Postman}, {Moustakas}, {Labbe},
  {Carrasco}, {Molino}, {Donahue}, {Kelson}, {Meneghetti}, {Jha}, {Benitez},
  {Lemze}, {Umetsu}, {Broadhurst}, {Moustakas}, {Rosati}, {Bartelmann}, {Ford},
  {Graves}, {Grillo}, {Infante}, {Jiminez-Teja}, {Jouvel}, {Lahav}, {Maoz},
  {Medezinski}, {Melchior}, {Merten}, {Nonino}, {Ogaz}, \& {Seitz}}]{bou12a}
{Bouwens}, R., {Bradley}, L., {Zitrin}, A., {et~al.} 2012{\natexlab{a}}, ArXiv
  e-prints

\bibitem[{{Bouwens} {et~al.}(2007){Bouwens}, {Illingworth}, {Franx}, \&
  {Ford}}]{bou07}
{Bouwens}, R.~J., {Illingworth}, G.~D., {Franx}, M., \& {Ford}, H. 2007, \apj,
  670, 928

\bibitem[{{Bouwens} {et~al.}(2009){Bouwens}, {Illingworth}, {Bradley}, {Ford},
  {Franx}, {Zheng}, {Broadhurst}, {Coe}, \& {Jee}}]{bou09}
{Bouwens}, R.~J., {Illingworth}, G.~D., {Bradley}, L.~D., {et~al.} 2009, \apj,
  690, 1764

\bibitem[{{Bouwens} {et~al.}(2011){Bouwens}, {Illingworth}, {Oesch},
  {Labb{\'e}}, {Trenti}, {van Dokkum}, {Franx}, {Stiavelli}, {Carollo},
  {Magee}, \& {Gonzalez}}]{bou11}
{Bouwens}, R.~J., {Illingworth}, G.~D., {Oesch}, P.~A., {et~al.} 2011, \apj,
  737, 90

\bibitem[{{Bouwens} {et~al.}(2012{\natexlab{b}}){Bouwens}, {Illingworth},
  {Oesch}, {Franx}, {Labb{\'e}}, {Trenti}, {van Dokkum}, {Carollo},
  {Gonz{\'a}lez}, {Smit}, \& {Magee}}]{bou12b}
---. 2012{\natexlab{b}}, \apj, 754, 83

\bibitem[{{Bovy} {et~al.}(2008){Bovy}, {Hogg}, \& {Moustakas}}]{bov08}
{Bovy}, J., {Hogg}, D.~W., \& {Moustakas}, J. 2008, \apj, 688, 198

\bibitem[{{Bradley} {et~al.}(2008){Bradley}, {Bouwens}, {Ford}, {Illingworth},
  {Jee}, {Ben{\'{\i}}tez}, {Broadhurst}, {Franx}, {Frye}, {Infante}, {Motta},
  {Rosati}, {White}, \& {Zheng}}]{bra08}
{Bradley}, L.~D., {Bouwens}, R.~J., {Ford}, H.~C., {et~al.} 2008, \apj, 678,
  647

\bibitem[{{Bradley} {et~al.}(2012){Bradley}, {Trenti}, {Oesch}, {Stiavelli},
  {Treu}, {Bouwens}, {Shull}, {Holwerda}, \& {Pirzkal}}]{bra12}
{Bradley}, L.~D., {Trenti}, M., {Oesch}, P.~A., {et~al.} 2012, \apj, 760, 108

\bibitem[{{Brammer} {et~al.}(2008){Brammer}, {van Dokkum}, \& {Coppi}}]{bram08}
{Brammer}, G.~B., {van Dokkum}, P.~G., \& {Coppi}, P. 2008, \apj, 686, 1503

\bibitem[{{Broadhurst} {et~al.}(1995){Broadhurst}, {Taylor}, \&
  {Peacock}}]{bro95}
{Broadhurst}, T.~J., {Taylor}, A.~N., \& {Peacock}, J.~A. 1995, \apj, 438, 49

\bibitem[{{Bruzual} \& {Charlot}(2003)}]{bru03}
{Bruzual}, G., \& {Charlot}, S. 2003, \mnras, 344, 1000

\bibitem[{{Bunker} {et~al.}(2004){Bunker}, {Stanway}, {Ellis}, \&
  {McMahon}}]{bun04}
{Bunker}, A.~J., {Stanway}, E.~R., {Ellis}, R.~S., \& {McMahon}, R.~G. 2004,
  \mnras, 355, 374

\bibitem[{{Bunker} {et~al.}(2010){Bunker}, {Wilkins}, {Ellis}, {Stark},
  {Lorenzoni}, {Chiu}, {Lacy}, {Jarvis}, \& {Hickey}}]{bun10}
{Bunker}, A.~J., {Wilkins}, S., {Ellis}, R.~S., {et~al.} 2010, \mnras, 409, 855

\bibitem[{{Calzetti} {et~al.}(2000){Calzetti}, {Armus}, {Bohlin}, {Kinney},
  {Koornneef}, \& {Storchi-Bergmann}}]{cal00}
{Calzetti}, D., {Armus}, L., {Bohlin}, R.~C., {et~al.} 2000, \apj, 533, 682

\bibitem[{{Calzetti} \& {Heckman}(1999)}]{cal99}
{Calzetti}, D., \& {Heckman}, T.~M. 1999, \apj, 519, 27

\bibitem[{{Calzetti} {et~al.}(1994){Calzetti}, {Kinney}, \&
  {Storchi-Bergmann}}]{cal94}
{Calzetti}, D., {Kinney}, A.~L., \& {Storchi-Bergmann}, T. 1994, \apj, 429, 582

\bibitem[{{Casertano} {et~al.}(2000){Casertano}, {de Mello}, {Dickinson},
  {Ferguson}, {Fruchter}, {Gonzalez-Lopezlira}, {Heyer}, {Hook}, {Levay},
  {Lucas}, {Mack}, {Makidon}, {Mutchler}, {Smith}, {Stiavelli}, {Wiggs}, \&
  {Williams}}]{cas00}
{Casertano}, S., {de Mello}, D., {Dickinson}, M., {et~al.} 2000, \aj, 120, 2747

\bibitem[{{Chabrier}(2003)}]{cha03}
{Chabrier}, G. 2003, \pasp, 115, 763

\bibitem[{{Charlot} \& {Fall}(2000)}]{cha00}
{Charlot}, S., \& {Fall}, S.~M. 2000, \apj, 539, 718

\bibitem[{{Chelouche} {et~al.}(2007){Chelouche}, {Koester}, \& {Bowen}}]{che07}
{Chelouche}, D., {Koester}, B.~P., \& {Bowen}, D.~V. 2007, \apjl, 671, L97

\bibitem[{{Coe} {et~al.}(2010){Coe}, {Ben{\'{\i}}tez}, {Broadhurst}, \&
  {Moustakas}}]{coe10}
{Coe}, D., {Ben{\'{\i}}tez}, N., {Broadhurst}, T., \& {Moustakas}, L.~A. 2010,
  \apj, 723, 1678

\bibitem[{{Coe} {et~al.}(2013){Coe}, {Zitrin}, {Carrasco}, {Shu}, {Zheng},
  {Postman}, {Bradley}, {Koekemoer}, {Bouwens}, {Broadhurst}, {Monna}, {Host},
  {Moustakas}, {Ford}, {Moustakas}, {van der Wel}, {Donahue}, {Rodney},
  {Ben{\'{\i}}tez}, {Jouvel}, {Seitz}, {Kelson}, \& {Rosati}}]{coe13}
{Coe}, D., {Zitrin}, A., {Carrasco}, M., {et~al.} 2013, \apj, 762, 32

\bibitem[{{Coleman} {et~al.}(1980){Coleman}, {Wu}, \& {Weedman}}]{col80}
{Coleman}, G.~D., {Wu}, C.-C., \& {Weedman}, D.~W. 1980, \apjs, 43, 393

\bibitem[{{Conroy} \& {Wechsler}(2009)}]{con09}
{Conroy}, C., \& {Wechsler}, R.~H. 2009, \apj, 696, 620

\bibitem[{{Conroy} {et~al.}(2006){Conroy}, {Wechsler}, \& {Kravtsov}}]{con06}
{Conroy}, C., {Wechsler}, R.~H., \& {Kravtsov}, A.~V. 2006, \apj, 647, 201

\bibitem[{{Daddi} {et~al.}(2004){Daddi}, {Cimatti}, {Renzini}, {Fontana},
  {Mignoli}, {Pozzetti}, {Tozzi}, \& {Zamorani}}]{dad04}
{Daddi}, E., {Cimatti}, A., {Renzini}, A., {et~al.} 2004, \apj, 617, 746

\bibitem[{{Egami} {et~al.}(2005){Egami}, {Kneib}, {Rieke}, {Ellis}, {Richard},
  {Rigby}, {Papovich}, {Stark}, {Santos}, {Huang}, {Dole}, {Le Floc'h}, \&
  {P{\'e}rez-Gonz{\'a}lez}}]{ega05}
{Egami}, E., {Kneib}, J.-P., {Rieke}, G.~H., {et~al.} 2005, \apjl, 618, L5

\bibitem[{{Erb} {et~al.}(2006){Erb}, {Shapley}, {Pettini}, {Steidel}, {Reddy},
  \& {Adelberger}}]{erb06}
{Erb}, D.~K., {Shapley}, A.~E., {Pettini}, M., {et~al.} 2006, \apj, 644, 813

\bibitem[{{Ferguson} {et~al.}(2000){Ferguson}, {Dickinson}, \&
  {Williams}}]{fer00}
{Ferguson}, H.~C., {Dickinson}, M., \& {Williams}, R. 2000, \araa, 38, 667

\bibitem[{{Fynbo} {et~al.}(2008){Fynbo}, {Prochaska}, {Sommer-Larsen},
  {Dessauges-Zavadsky}, \& {M{\o}ller}}]{fyn08}
{Fynbo}, J.~P.~U., {Prochaska}, J.~X., {Sommer-Larsen}, J.,
  {Dessauges-Zavadsky}, M., \& {M{\o}ller}, P. 2008, \apj, 683, 321

\bibitem[{{Governato} {et~al.}(2012){Governato}, {Zolotov}, {Pontzen},
  {Christensen}, {Oh}, {Brooks}, {Quinn}, {Shen}, \& {Wadsley}}]{gov12}
{Governato}, F., {Zolotov}, A., {Pontzen}, A., {et~al.} 2012, \mnras, 422, 1231

\bibitem[{{Haberzettl} {et~al.}(2012){Haberzettl}, {Williger}, {Lehnert},
  {Nesvadba}, \& {Davies}}]{hab12}
{Haberzettl}, L., {Williger}, G., {Lehnert}, M.~D., {Nesvadba}, N., \&
  {Davies}, L. 2012, \apj, 745, 96

\bibitem[{{Hathi} {et~al.}(2010){Hathi}, {Ryan}, {Cohen}, {Yan}, {Windhorst},
  {McCarthy}, {O'Connell}, {Koekemoer}, {Rutkowski}, {Balick}, {Bond},
  {Calzetti}, {Disney}, {Dopita}, {Frogel}, {Hall}, {Holtzman}, {Kimble},
  {Paresce}, {Saha}, {Silk}, {Trauger}, {Walker}, {Whitmore}, \&
  {Young}}]{hat10}
{Hathi}, N.~P., {Ryan}, Jr., R.~E., {Cohen}, S.~H., {et~al.} 2010, \apj, 720,
  1708

\bibitem[{{Hathi} {et~al.}(2012){Hathi}, {Cohen}, {Ryan}, {Finkelstein},
  {McCarthy}, {Windhorst}, {Yan}, {Koekemoer}, {Rutkowski}, {O'Connell},
  {Straughn}, {Balick}, {Bond}, {Calzetti}, {Disney}, {Dopita}, {Frogel},
  {Hall}, {Holtzman}, {Kimble}, {Paresce}, {Saha}, {Silk}, {Trauger}, {Walker},
  {Whitmore}, \& {Young}}]{hat12}
{Hathi}, N.~P., {Cohen}, S.~H., {Ryan}, Jr., R.~E., {et~al.} 2012, ArXiv
  e-prints

\bibitem[{{Hathi} {et~al.}(2013){Hathi}, {Cohen}, {Ryan}, {Finkelstein},
  {McCarthy}, {Windhorst}, {Yan}, {Koekemoer}, {Rutkowski}, {O'Connell},
  {Straughn}, {Balick}, {Bond}, {Calzetti}, {Disney}, {Dopita}, {Frogel},
  {Hall}, {Holtzman}, {Kimble}, {Paresce}, {Saha}, {Silk}, {Trauger}, {Walker},
  {Whitmore}, \& {Young}}]{hat13}
---. 2013, \apj, 765, 88

\bibitem[{{Hopkins} \& {Beacom}(2006)}]{hop06}
{Hopkins}, A.~M., \& {Beacom}, J.~F. 2006, \apj, 651, 142

\bibitem[{{Hopkins} {et~al.}(2000){Hopkins}, {Connolly}, \& {Szalay}}]{hop00}
{Hopkins}, A.~M., {Connolly}, A.~J., \& {Szalay}, A.~S. 2000, \aj, 120, 2843

\bibitem[{{Iwata} {et~al.}(2007){Iwata}, {Ohta}, {Tamura}, {Akiyama}, {Aoki},
  {Ando}, {Kiuchi}, \& {Sawicki}}]{iwa07}
{Iwata}, I., {Ohta}, K., {Tamura}, N., {et~al.} 2007, \mnras, 376, 1557

\bibitem[{{Janknecht} {et~al.}(2006){Janknecht}, {Reimers}, {Lopez}, \&
  {Tytler}}]{jan06}
{Janknecht}, E., {Reimers}, D., {Lopez}, S., \& {Tytler}, D. 2006, \aap, 458,
  427

\bibitem[{{Jones} {et~al.}(2010){Jones}, {Ellis}, {Jullo}, \&
  {Richard}}]{jon10}
{Jones}, T., {Ellis}, R., {Jullo}, E., \& {Richard}, J. 2010, \apjl, 725, L176

\bibitem[{{Jullo} {et~al.}(2007){Jullo}, {Kneib}, {Limousin},
  {El{\'{\i}}asd{\'o}ttir}, {Marshall}, \& {Verdugo}}]{jul07}
{Jullo}, E., {Kneib}, J.-P., {Limousin}, M., {et~al.} 2007, New Journal of
  Physics, 9, 447

\bibitem[{{Kennicutt}(1998)}]{ken98}
{Kennicutt}, Jr., R.~C. 1998, \araa, 36, 189

\bibitem[{{Kim} {et~al.}(1997){Kim}, {Hu}, {Cowie}, \& {Songaila}}]{kim97}
{Kim}, T.-S., {Hu}, E.~M., {Cowie}, L.~L., \& {Songaila}, A. 1997, \aj, 114, 1

\bibitem[{{Kneib} {et~al.}(2004){Kneib}, {Ellis}, {Santos}, \&
  {Richard}}]{kne04}
{Kneib}, J.-P., {Ellis}, R.~S., {Santos}, M.~R., \& {Richard}, J. 2004, \apj,
  607, 697

\bibitem[{{Koekemoer} {et~al.}(2003){Koekemoer}, {Fruchter}, {Hook}, \&
  {Hack}}]{koe03}
{Koekemoer}, A.~M., {Fruchter}, A.~S., {Hook}, R.~N., \& {Hack}, W. 2003, in
  HST Calibration Workshop : Hubble after the Installation of the ACS and the
  NICMOS Cooling System, ed. S.~{Arribas}, A.~{Koekemoer}, \& B.~{Whitmore},
  337

\bibitem[{{Kroupa}(2001)}]{kro01}
{Kroupa}, P. 2001, \mnras, 322, 231

\bibitem[{{Kuhlen} \& {Faucher-Gigu{\`e}re}(2012)}]{kuh12}
{Kuhlen}, M., \& {Faucher-Gigu{\`e}re}, C.-A. 2012, \mnras, 423, 862

\bibitem[{{Kuhlen} {et~al.}(2013){Kuhlen}, {Madau}, \& {Krumholz}}]{kuh13}
{Kuhlen}, M., {Madau}, P., \& {Krumholz}, M. 2013, ArXiv e-prints

\bibitem[{{Labb{\'e}} {et~al.}(2007){Labb{\'e}}, {Franx}, {Rudnick},
  {Schreiber}, {van Dokkum}, {Moorwood}, {Rix}, {R{\"o}ttgering}, {Trujillo},
  \& {van der Werf}}]{lab07}
{Labb{\'e}}, I., {Franx}, M., {Rudnick}, G., {et~al.} 2007, \apj, 665, 944

\bibitem[{{Law} {et~al.}(2012){Law}, {Steidel}, {Shapley}, {Nagy}, {Reddy}, \&
  {Erb}}]{law12}
{Law}, D.~R., {Steidel}, C.~C., {Shapley}, A.~E., {et~al.} 2012, ArXiv e-prints

\bibitem[{{Leitherer} {et~al.}(1999){Leitherer}, {Schaerer}, {Goldader},
  {Gonz{\'a}lez Delgado}, {Robert}, {Kune}, {de Mello}, {Devost}, \&
  {Heckman}}]{lei99}
{Leitherer}, C., {Schaerer}, D., {Goldader}, J.~D., {et~al.} 1999, \apjs, 123,
  3

\bibitem[{{Lilly} {et~al.}(1996){Lilly}, {Le Fevre}, {Hammer}, \&
  {Crampton}}]{lil96}
{Lilly}, S.~J., {Le Fevre}, O., {Hammer}, F., \& {Crampton}, D. 1996, \apjl,
  460, L1

\bibitem[{{Limousin} {et~al.}(2007){Limousin}, {Richard}, {Jullo}, {Kneib},
  {Fort}, {Soucail}, {El{\'{\i}}asd{\'o}ttir}, {Natarajan}, {Ellis}, {Smail},
  {Czoske}, {Smith}, {Hudelot}, {Bardeau}, {Ebeling}, {Egami}, \&
  {Knudsen}}]{lim07}
{Limousin}, M., {Richard}, J., {Jullo}, E., {et~al.} 2007, \apj, 668, 643

\bibitem[{{Ly} {et~al.}(2009){Ly}, {Malkan}, {Treu}, {Woo}, {Currie},
  {Hayashi}, {Kashikawa}, {Motohara}, {Shimasaku}, \& {Yoshida}}]{ly09}
{Ly}, C., {Malkan}, M.~A., {Treu}, T., {et~al.} 2009, \apj, 697, 1410

\bibitem[{{Madau}(1995)}]{mad95}
{Madau}, P. 1995, \apj, 441, 18

\bibitem[{{Madau} {et~al.}(1996){Madau}, {Ferguson}, {Dickinson}, {Giavalisco},
  {Steidel}, \& {Fruchter}}]{mad96}
{Madau}, P., {Ferguson}, H.~C., {Dickinson}, M.~E., {et~al.} 1996, \mnras, 283,
  1388

\bibitem[{{Magnelli} {et~al.}(2011){Magnelli}, {Elbaz}, {Chary}, {Dickinson},
  {Le Borgne}, {Frayer}, \& {Willmer}}]{mag11}
{Magnelli}, B., {Elbaz}, D., {Chary}, R.~R., {et~al.} 2011, \aap, 528, A35

\bibitem[{{McLure} {et~al.}(2009){McLure}, {Cirasuolo}, {Dunlop}, {Foucaud}, \&
  {Almaini}}]{mcl09}
{McLure}, R.~J., {Cirasuolo}, M., {Dunlop}, J.~S., {Foucaud}, S., \& {Almaini},
  O. 2009, \mnras, 395, 2196

\bibitem[{{Meurer} {et~al.}(1999){Meurer}, {Heckman}, \& {Calzetti}}]{meu99}
{Meurer}, G.~R., {Heckman}, T.~M., \& {Calzetti}, D. 1999, \apj, 521, 64

\bibitem[{{Moller} \& {Jakobsen}(1990)}]{mol90}
{Moller}, P., \& {Jakobsen}, P. 1990, \aap, 228, 299

\bibitem[{{Muller} {et~al.}(2008){Muller}, {Wu}, {Hsieh}, {Gonz{\'a}lez},
  {Loinard}, {Yee}, \& {Gladders}}]{mul08}
{Muller}, S., {Wu}, S.-Y., {Hsieh}, B.-C., {et~al.} 2008, \apj, 680, 975

\bibitem[{{Narayan} {et~al.}(1984){Narayan}, {Blandford}, \&
  {Nityananda}}]{nar84}
{Narayan}, R., {Blandford}, R., \& {Nityananda}, R. 1984, \nat, 310, 112

\bibitem[{{Nestor} {et~al.}(2012){Nestor}, {Shapley}, {Kornei}, {Steidel}, \&
  {Siana}}]{nes12}
{Nestor}, D.~B., {Shapley}, A.~E., {Kornei}, K.~A., {Steidel}, C.~C., \&
  {Siana}, B. 2012, ArXiv e-prints

\bibitem[{{Nestor} {et~al.}(2011){Nestor}, {Shapley}, {Steidel}, \&
  {Siana}}]{nes11}
{Nestor}, D.~B., {Shapley}, A.~E., {Steidel}, C.~C., \& {Siana}, B. 2011, \apj,
  736, 18

\bibitem[{{Noeske} {et~al.}(2012){Noeske}, {Baggett}, {Bushouse}, {Petro},
  {Gilliland}, \& {Khozurina-Platais}}]{noe12}
{Noeske}, K., {Baggett}, S., {Bushouse}, H., {et~al.} 2012, {WFC3 UVIS Charge
  Transfer Eciency October 2009 to October 2011}, Tech. rep.

\bibitem[{{Oesch} {et~al.}(2007){Oesch}, {Stiavelli}, {Carollo}, {Bergeron},
  {Koekemoer}, {Lucas}, {Pavlovsky}, {Trenti}, {Lilly}, {Beckwith}, {Dahlen},
  {Ferguson}, {Gardner}, {Lacey}, {Mobasher}, {Panagia}, \& {Rix}}]{oes07}
{Oesch}, P.~A., {Stiavelli}, M., {Carollo}, C.~M., {et~al.} 2007, \apj, 671,
  1212

\bibitem[{{Oesch} {et~al.}(2010a){Oesch}, {Bouwens}, {Carollo}, {Illingworth},
  {Magee}, {Trenti}, {Stiavelli}, {Franx}, {Labb{\'e}}, \& {van
  Dokkum}}]{oes10a}
{Oesch}, P.~A., {Bouwens}, R.~J., {Carollo}, C.~M., {et~al.} 2010a, \apjl, 725,
  L150

\bibitem[{{Oesch} {et~al.}(2010b){Oesch}, {Bouwens}, {Illingworth}, {Carollo},
  {Franx}, {Labb{\'e}}, {Magee}, {Stiavelli}, {Trenti}, \& {van
  Dokkum}}]{oes10b}
{Oesch}, P.~A., {Bouwens}, R.~J., {Illingworth}, G.~D., {et~al.} 2010b, \apjl,
  709, L16

\bibitem[{{Oke} \& {Gunn}(1983)}]{oke83}
{Oke}, J.~B., \& {Gunn}, J.~E. 1983, \apj, 266, 713

\bibitem[{{O'Meara} {et~al.}(2012){O'Meara}, {Prochaska}, {Worseck}, {Chen}, \&
  {Madau}}]{ome12}
{O'Meara}, J.~M., {Prochaska}, J.~X., {Worseck}, G., {Chen}, H.-W., \& {Madau},
  P. 2012, ArXiv e-prints

\bibitem[{{Ouchi} {et~al.}(2009){Ouchi}, {Mobasher}, {Shimasaku}, {Ferguson},
  {Fall}, {Ono}, {Kashikawa}, {Morokuma}, {Nakajima}, {Okamura}, {Dickinson},
  {Giavalisco}, \& {Ohta}}]{ouc09}
{Ouchi}, M., {Mobasher}, B., {Shimasaku}, K., {et~al.} 2009, \apj, 706, 1136

\bibitem[{{Overzier} {et~al.}(2008){Overzier}, {Bouwens}, {Cross}, {Venemans},
  {Miley}, {Zirm}, {Ben{\'{\i}}tez}, {Blakeslee}, {Coe}, {Demarco}, {Ford},
  {Homeier}, {Illingworth}, {Kurk}, {Martel}, {Mei}, {Oliveira},
  {R{\"o}ttgering}, {Tsvetanov}, \& {Zheng}}]{ove08}
{Overzier}, R.~A., {Bouwens}, R.~J., {Cross}, N.~J.~G., {et~al.} 2008, \apj,
  673, 143

\bibitem[{{Pettini} {et~al.}(2002){Pettini}, {Rix}, {Steidel}, {Hunt},
  {Shapley}, \& {Adelberger}}]{pet02}
{Pettini}, M., {Rix}, S.~A., {Steidel}, C.~C., {et~al.} 2002, \apss, 281, 461

\bibitem[{{Pickles}(1998)}]{pic98}
{Pickles}, A.~J. 1998, VizieR Online Data Catalog, 611, 863

\bibitem[{{Rao} {et~al.}(2006){Rao}, {Turnshek}, \& {Nestor}}]{rao06}
{Rao}, S.~M., {Turnshek}, D.~A., \& {Nestor}, D.~B. 2006, \apj, 636, 610

\bibitem[{{Reddy} {et~al.}(2010){Reddy}, {Erb}, {Pettini}, {Steidel}, \&
  {Shapley}}]{red10}
{Reddy}, N.~A., {Erb}, D.~K., {Pettini}, M., {Steidel}, C.~C., \& {Shapley},
  A.~E. 2010, \apj, 712, 1070

\bibitem[{{Reddy} \& {Steidel}(2009)}]{red09}
{Reddy}, N.~A., \& {Steidel}, C.~C. 2009, \apj, 692, 778

\bibitem[{{Reddy} {et~al.}(2006){Reddy}, {Steidel}, {Fadda}, {Yan}, {Pettini},
  {Shapley}, {Erb}, \& {Adelberger}}]{red06}
{Reddy}, N.~A., {Steidel}, C.~C., {Fadda}, D., {et~al.} 2006, \apj, 644, 792

\bibitem[{{Reddy} {et~al.}(2008){Reddy}, {Steidel}, {Pettini}, {Adelberger},
  {Shapley}, {Erb}, \& {Dickinson}}]{red08}
{Reddy}, N.~A., {Steidel}, C.~C., {Pettini}, M., {et~al.} 2008, \apjs, 175, 48

\bibitem[{{Ribaudo} {et~al.}(2011){Ribaudo}, {Lehner}, \& {Howk}}]{rib11}
{Ribaudo}, J., {Lehner}, N., \& {Howk}, J.~C. 2011, \apj, 736, 42

\bibitem[{{Richard} {et~al.}(2011){Richard}, {Kneib}, {Ebeling}, {Stark},
  {Egami}, \& {Fiedler}}]{ric11}
{Richard}, J., {Kneib}, J.-P., {Ebeling}, H., {et~al.} 2011, \mnras, 414, L31

\bibitem[{{Richard} {et~al.}(2008){Richard}, {Stark}, {Ellis}, {George},
  {Egami}, {Kneib}, \& {Smith}}]{ric08}
{Richard}, J., {Stark}, D.~P., {Ellis}, R.~S., {et~al.} 2008, \apj, 685, 705

\bibitem[{{Rix} {et~al.}(2004){Rix}, {Barden}, {Beckwith}, {Bell}, {Borch},
  {Caldwell}, {H{\"a}ussler}, {Jahnke}, {Jogee}, {McIntosh}, {Meisenheimer},
  {Peng}, {Sanchez}, {Somerville}, {Wisotzki}, \& {Wolf}}]{rix04}
{Rix}, H.-W., {Barden}, M., {Beckwith}, S.~V.~W., {et~al.} 2004, \apjs, 152,
  163

\bibitem[{{Robertson}(2010)}]{rob10}
{Robertson}, B.~E. 2010, \apj, 713, 1266

\bibitem[{{Robertson} {et~al.}(2013){Robertson}, {Furlanetto}, {Schneider},
  {Charlot}, {Ellis}, {Stark}, {McLure}, {Dunlop}, {Koekemoer}, {Schenker},
  {Ouchi}, {Ono}, {Curtis-Lake}, {Rogers}, {Bowler}, \& {Cirasuolo}}]{rob13}
{Robertson}, B.~E., {Furlanetto}, S.~R., {Schneider}, E., {et~al.} 2013, ArXiv
  e-prints

\bibitem[{{Salpeter}(1955)}]{sal55}
{Salpeter}, E.~E. 1955, \apj, 121, 161

\bibitem[{{Sandage} {et~al.}(1979){Sandage}, {Tammann}, \& {Yahil}}]{san79}
{Sandage}, A., {Tammann}, G.~A., \& {Yahil}, A. 1979, \apj, 232, 352

\bibitem[{{Sawicki}(2012)}]{saw12}
{Sawicki}, M. 2012, \mnras, 421, 2187

\bibitem[{{Sawicki} \& {Thompson}(2006)}]{saw06}
{Sawicki}, M., \& {Thompson}, D. 2006, \apj, 642, 653

\bibitem[{{Schenker} {et~al.}(2012){Schenker}, {Robertson}, {Ellis}, {Ono},
  {McLure}, {Dunlop}, {Koekemoer}, {Bowler}, {Ouchi}, {Curtis-Lake}, {Rogers},
  {Schneider}, {Charlot}, {Stark}, {Furlanetto}, \& {Cirasuolo}}]{sch12}
{Schenker}, M.~A., {Robertson}, B.~E., {Ellis}, R.~S., {et~al.} 2012, ArXiv
  e-prints

\bibitem[{{Shapley} {et~al.}(2005){Shapley}, {Steidel}, {Erb}, {Reddy},
  {Adelberger}, {Pettini}, {Barmby}, \& {Huang}}]{sha05}
{Shapley}, A.~E., {Steidel}, C.~C., {Erb}, D.~K., {et~al.} 2005, \apj, 626, 698

\bibitem[{{Siana} {et~al.}(2008{\natexlab{a}}){Siana}, {Teplitz}, {Chary},
  {Colbert}, \& {Frayer}}]{sia08b}
{Siana}, B., {Teplitz}, H.~I., {Chary}, R.-R., {Colbert}, J., \& {Frayer},
  D.~T. 2008{\natexlab{a}}, \apj, 689, 59

\bibitem[{{Siana} {et~al.}(2008{\natexlab{b}}){Siana}, {Polletta}, {Smith},
  {Lonsdale}, {Gonzalez-Solares}, {Farrah}, {Babbedge}, {Rowan-Robinson},
  {Surace}, {Shupe}, {Fang}, {Franceschini}, \& {Oliver}}]{sia08a}
{Siana}, B., {Polletta}, M.~d.~C., {Smith}, H.~E., {et~al.} 2008{\natexlab{b}},
  \apj, 675, 49

\bibitem[{{Siana} {et~al.}(2009){Siana}, {Smail}, {Swinbank}, {Richard},
  {Teplitz}, {Coppin}, {Ellis}, {Stark}, {Kneib}, \& {Edge}}]{sia09}
{Siana}, B., {Smail}, I., {Swinbank}, A.~M., {et~al.} 2009, \apj, 698, 1273

\bibitem[{{Stark} {et~al.}(2007){Stark}, {Ellis}, {Richard}, {Kneib}, {Smith},
  \& {Santos}}]{sta07}
{Stark}, D.~P., {Ellis}, R.~S., {Richard}, J., {et~al.} 2007, \apj, 663, 10

\bibitem[{{Stark} {et~al.}(2008){Stark}, {Swinbank}, {Ellis}, {Dye}, {Smail},
  \& {Richard}}]{sta08}
{Stark}, D.~P., {Swinbank}, A.~M., {Ellis}, R.~S., {et~al.} 2008, \nat, 455,
  775

\bibitem[{{Steidel} {et~al.}(1999){Steidel}, {Adelberger}, {Giavalisco},
  {Dickinson}, \& {Pettini}}]{ste99}
{Steidel}, C.~C., {Adelberger}, K.~L., {Giavalisco}, M., {Dickinson}, M., \&
  {Pettini}, M. 1999, \apj, 519, 1

\bibitem[{{Teplitz} {et~al.}(2013){Teplitz}, {Rafelski}, {Kurczynski}, {Bond},
  {Grogin}, {Koekemoer}, {Atek}, {Brown}, {Coe}, {Colbert}, {Ferguson},
  {Finkelstein}, {Gardner}, {Gawiser}, {Giavalisco}, {Gronwall}, {Hanish},
  {Lee}, {de Mello}, {Ravindranath}, {Ryan}, {Siana}, {Scarlata}, {Soto},
  {Voyer}, \& {Wolfe}}]{tep13}
{Teplitz}, H.~I., {Rafelski}, M., {Kurczynski}, P., {et~al.} 2013, ArXiv
  e-prints

\bibitem[{{Tody}(1986)}]{tod86}
{Tody}, D. 1986, in Society of Photo-Optical Instrumentation Engineers (SPIE)
  Conference Series, Vol. 627, Society of Photo-Optical Instrumentation
  Engineers (SPIE) Conference Series, ed. D.~L. {Crawford}, 733

\bibitem[{{Wilkins} {et~al.}(2011){Wilkins}, {Bunker}, {Stanway}, {Lorenzoni},
  \& {Caruana}}]{wil11}
{Wilkins}, S.~M., {Bunker}, A.~J., {Stanway}, E., {Lorenzoni}, S., \&
  {Caruana}, J. 2011, \mnras, 417, 717

\bibitem[{{Windhorst} {et~al.}(2011){Windhorst}, {Cohen}, {Hathi}, {McCarthy},
  {Ryan}, {Yan}, {Baldry}, {Driver}, {Frogel}, {Hill}, {Kelvin}, {Koekemoer},
  {Mechtley}, {O'Connell}, {Robotham}, {Rutkowski}, {Seibert}, {Straughn},
  {Tuffs}, {Balick}, {Bond}, {Bushouse}, {Calzetti}, {Crockett}, {Disney},
  {Dopita}, {Hall}, {Holtzman}, {Kaviraj}, {Kimble}, {MacKenty}, {Mutchler},
  {Paresce}, {Saha}, {Silk}, {Trauger}, {Walker}, {Whitmore}, \&
  {Young}}]{win11}
{Windhorst}, R.~A., {Cohen}, S.~H., {Hathi}, N.~P., {et~al.} 2011, \apjs, 193,
  27

\bibitem[{{Wyder} {et~al.}(2005){Wyder}, {Treyer}, {Milliard}, {Schiminovich},
  {Arnouts}, {Budav{\'a}ri}, {Barlow}, {Bianchi}, {Byun}, {Donas}, {Forster},
  {Friedman}, {Heckman}, {Jelinsky}, {Lee}, {Madore}, {Malina}, {Martin},
  {Morrissey}, {Neff}, {Rich}, {Siegmund}, {Small}, {Szalay}, \&
  {Welsh}}]{wyd05}
{Wyder}, T.~K., {Treyer}, M.~A., {Milliard}, B., {et~al.} 2005, \apjl, 619, L15

\bibitem[{{Yan} {et~al.}(2006){Yan}, {Dickinson}, {Giavalisco}, {Stern},
  {Eisenhardt}, \& {Ferguson}}]{yan06}
{Yan}, H., {Dickinson}, M., {Giavalisco}, M., {et~al.} 2006, \apj, 651, 24

\bibitem[{{Yan} {et~al.}(2010){Yan}, {Windhorst}, {Hathi}, {Cohen}, {Ryan},
  {O'Connell}, \& {McCarthy}}]{yan10}
{Yan}, H.-J., {Windhorst}, R.~A., {Hathi}, N.~P., {et~al.} 2010, Research in
  Astronomy and Astrophysics, 10, 867

\bibitem[{{Yoshida} {et~al.}(2006){Yoshida}, {Shimasaku}, {Kashikawa}, {Ouchi},
  {Okamura}, {Ajiki}, {Akiyama}, {Ando}, {Aoki}, {Doi}, {Furusawa},
  {Hayashino}, {Iwamuro}, {Iye}, {Karoji}, {Kobayashi}, {Kodaira}, {Kodama},
  {Komiyama}, {Malkan}, {Matsuda}, {Miyazaki}, {Mizumoto}, {Morokuma},
  {Motohara}, {Murayama}, {Nagao}, {Nariai}, {Ohta}, {Sasaki}, {Sato},
  {Sekiguchi}, {Shioya}, {Tamura}, {Taniguchi}, {Umemura}, {Yamada}, \&
  {Yasuda}}]{yos06}
{Yoshida}, M., {Shimasaku}, K., {Kashikawa}, N., {et~al.} 2006, \apj, 653, 988

\bibitem[{{Yuan} {et~al.}(2013){Yuan}, {Kewley}, \& {Richard}}]{yua13}
{Yuan}, T.-T., {Kewley}, L.~J., \& {Richard}, J. 2013, \apj, 763, 9

\bibitem[{{Zheng} {et~al.}(2012){Zheng}, {Postman}, {Zitrin}, {Moustakas},
  {Shu}, {Jouvel}, {H{\o}st}, {Molino}, {Bradley}, {Coe}, {Moustakas},
  {Carrasco}, {Ford}, {Ben{\'{\i}}tez}, {Lauer}, {Seitz}, {Bouwens},
  {Koekemoer}, {Medezinski}, {Bartelmann}, {Broadhurst}, {Donahue}, {Grillo},
  {Infante}, {Jha}, {Kelson}, {Lahav}, {Lemze}, {Melchior}, {Meneghetti},
  {Merten}, {Nonino}, {Ogaz}, {Rosati}, {Umetsu}, \& {van der Wel}}]{zhe12}
{Zheng}, W., {Postman}, M., {Zitrin}, A., {et~al.} 2012, \nat, 489, 406

\bibitem[{{Zitrin} {et~al.}(2012){Zitrin}, {Moustakas}, {Bradley}, {Coe},
  {Moustakas}, {Postman}, {Shu}, {Zheng}, {Ben{\'{\i}}tez}, {Bouwens},
  {Broadhurst}, {Ford}, {Host}, {Jouvel}, {Koekemoer}, {Meneghetti}, {Rosati},
  {Donahue}, {Grillo}, {Kelson}, {Lemze}, {Medezinski}, {Molino}, {Nonino}, \&
  {Ogaz}}]{zit12}
{Zitrin}, A., {Moustakas}, J., {Bradley}, L., {et~al.} 2012, \apjl, 747, L9

\end{thebibliography}
\clearpage

\end{document}